\documentclass[10pt,twocolumn,letterpaper]{article}

\usepackage{cvpr}
\usepackage{times}
\usepackage{epsfig}
\usepackage{graphicx}
\usepackage{amsmath}
\usepackage{amssymb}
\usepackage{subcaption}

\usepackage[breaklinks=true,bookmarks=false]{hyperref}

\cvprfinalcopy 

\def\cvprPaperID{89} 

\ifcvprfinal\fi

\newcommand{\specialcell}[2][c]{%
  \begin{tabular}[#1]{@{}c@{}}#2\end{tabular}}

\begin{document}

\title{Image Colorization By Capsule Networks}

\author{G\"{o}khan \"{O}zbulak$^{1,2}$ \\
$^1$Istanbul Technical University\\ 
$^2$The Scientific And Technological Research Council Of Turkey\\
{\tt\small gokhan.ozbulak@\{itu.edu.tr,tubitak.gov.tr\}}
}

\maketitle
\thispagestyle{empty}

\begin{abstract}
   In this paper, a simple topology of Capsule Network (CapsNet) is investigated for the 
   problem of image colorization. The generative and segmentation capabilities of the 
   original CapsNet topology, which is proposed for image classification problem, is 
   leveraged for the colorization of the images by modifying the network as follows: 
   1) The original CapsNet model is adapted to map the grayscale input to the output in 
   the CIE Lab colorspace, 2) The feature detector part of the model is updated by using 
   deeper feature layers inherited from VGG-19 pre-trained model with weights in order to 
   transfer low-level image representation capability to this model, 3) The margin loss 
   function is modified as Mean Squared Error (MSE) loss to minimize the image-to-image 
   mapping. The resulting CapsNet model is named as Colorizer Capsule Network (ColorCapsNet). 
   The performance of the ColorCapsNet is evaluated on the DIV2K dataset and promising results 
   are obtained to investigate Capsule Networks further for image colorization problem.
\end{abstract}

\section{Introduction}

Image colorization is the problem of converting the image from grayscale to the another 
colorspace so that the image is colorized. As the colorization problem requires a mapping 
from the simpler data (one-channel grayscale image) to the more complex data (multi-channel 
composite image), many different mappings may be obtained with the most of them is far from 
satisfactory colorization. Hence, the problem of image colorization is stated as ill-posed.

\begin{figure}[h!]
  \centering
  \begin{subfigure}[b]{0.4\linewidth}
    \includegraphics[width=\linewidth]{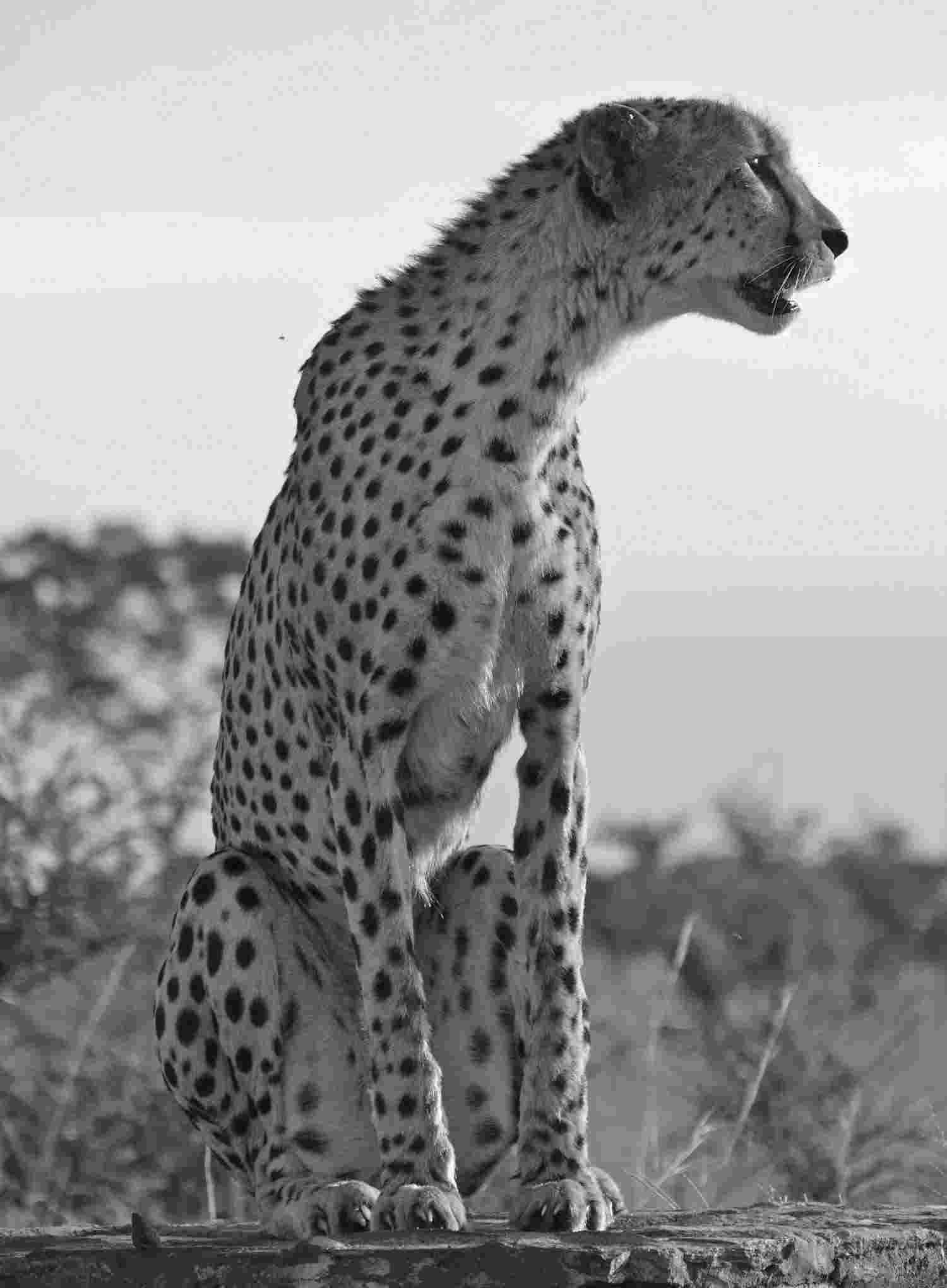}
  \end{subfigure}
  \begin{subfigure}[b]{0.4\linewidth}
    \includegraphics[width=\linewidth]{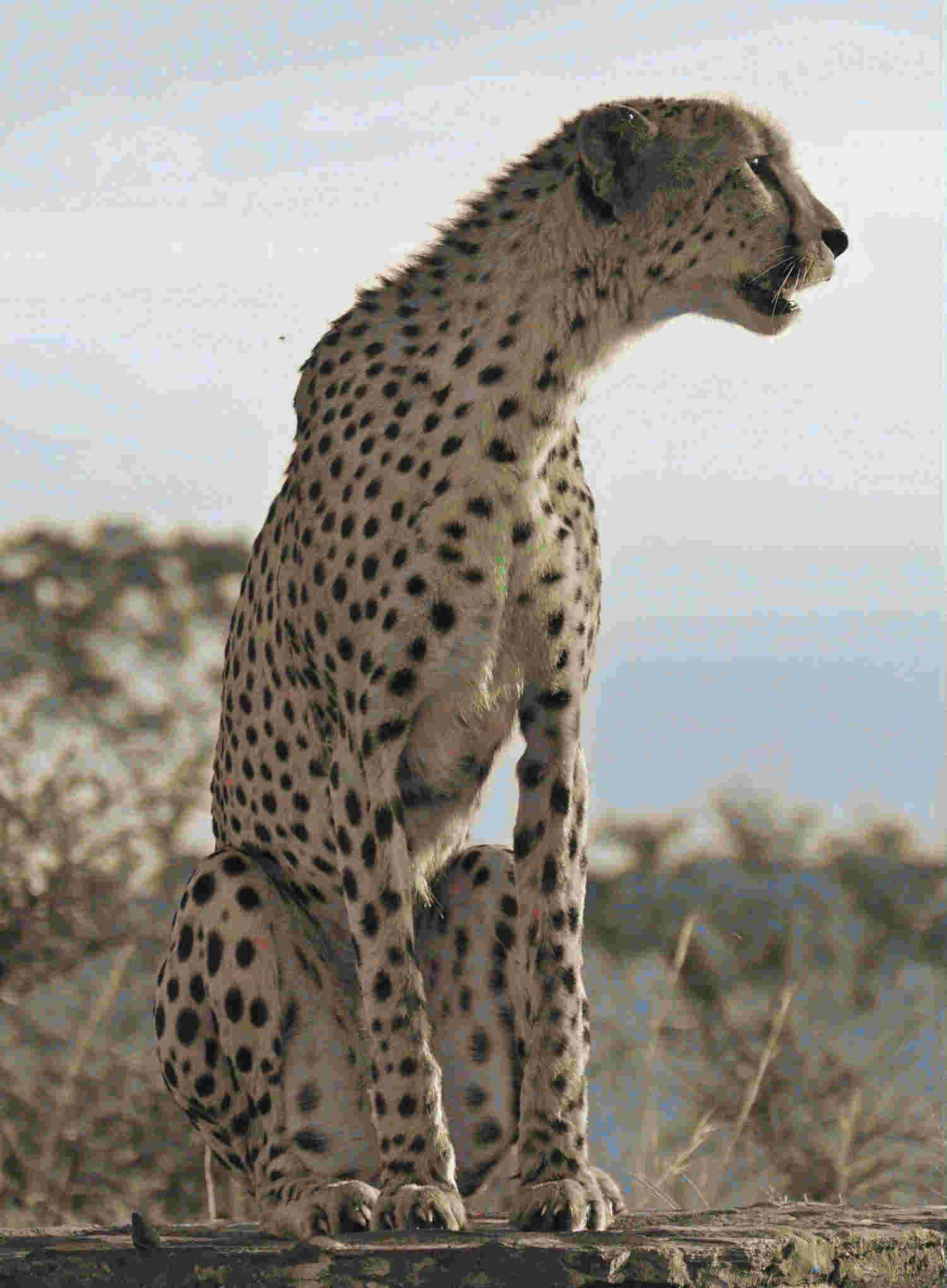}
  \end{subfigure}
  \caption{An example of colorization by the ColorCapsNet (psnr=23.74,ssim=0.91).}
  \label{fig:highlight}
\end{figure}

There are many methods proposed in the literature to tackle down the ill-posed nature of the 
colorization problem,  These studies may be classified as in two categories: 
Colorization with 1) guidance \cite{chia2011semantic,gupta2012image,ironi2005colorization,
levin2004colorization,yatziv2006fast}, and 2) no guidance \cite{charpiat2008automatic,cheng2015deep,
iizuka2016let,larsson2016learning,zhang2016colorful}. In guided colorization, 
user interaction/example image is asked to provide feedback about colorization and this 
feedback is combined with the colorization algorithm to obtain satisfactory results. In 
colorization with no guidance, automatic colorization algorithms are considered with excluding 
user intervention. The latter approach is harder to get satisfactory colorization results 
because fully automatic methods may fail to decide for the proper colors when the alternative colors 
are possible for the object in interest.

In recent years, with the advancements in deep learning, many convolutional and generative 
deep models have arised to tackle down the challenging image analysis problems varying from 
object classification and detection to enhancement (denoising, colorization etc.). 
Especially, the problem of automatic image colorization is massively investigated by 
leveraging successful deep models and promising results are obtained quantitatively and 
qualitatively with these models.

A recent study, which proposes a deep architecture named as Capsule Network (CapsNet) 
\cite{sabour2017dynamic}, introduces a network capable both of image classification and 
generation. The experimental results of proposed method in \cite{sabour2017dynamic} show 
that state-of-the-art (SOTA) performance may be obtained on image classification task by 
using a shallow CapsNet architecture. On the other hand, the image generation/latent space representation 
capability remains an open research area for further investigation.

In this paper, the generative characteristic of the CapsNet model is considered for the 
image colorization problem. The original architecture proposed in \cite{sabour2017dynamic} 
consists of one feature detection layer (convolutional layer), one feature representation 
layer (primary capsule layer) and one classification layer (capsule layer), and is trained 
for the digit classification task. Here, first, the capacity of the feature detector is increased 
by adding more convolutional layers. The resulting feature detector is same as the first 
two convolutional layers of VGG-19 model \cite{simonyan2014very}. For these layers, the 
weights from VGG-19 model pre-trained on ILSVRC 2012 \cite{deng2009imagenet} are also 
transferred before training in order to initialize the network with prior low-level feature 
representation. This may be regarded as a transfer learning strategy. Then, the network is 
adapted to generate the image in CIE Lab colospace from its grayscale counterpart. Finally, 
the margin loss, which is defined for the classification task, is changed to Mean Squared Error 
(MSE) loss in order to minimize the difference between real and generated color images. 
The resulting colorization model is named as Colorizer Capsule Network (ColorCapsNet). 
A colorization example by ColorCapsNet can be examined in Figure~\ref{fig:highlight}.

ColorCapsNet is trained on two different datasets. First, it's trained on ILSVRC 2012 
dataset \cite{deng2009imagenet} in order to learn the general color distribution of the 
objects. Then, DIV2K dataset \cite{agustsson2017ntire} is used to obtain final colorization 
model. In both cases, the datasets are pre-processed for proper colorspace conversion and 
data size. Proposed method is patch-based so that a pre-defined patch 
size must be determined before training. The patch-based mapping exploits the information in 
local neighborhood so that the pixel in interest is colorized according to the color 
distribution in its neighborhood \cite{charpiat2008automatic}.

The organization of the paper is as follows: In Section \ref{relatedwork}, the literature 
is reviewed for both image colorization with guidance and no guidance. In 
Section \ref{proposedmethod}, the ColorCapsNet model is explained in detail. 
The performance of the method is discussed in Section \ref{experimentalanalysis} and the 
paper is concluded in Section \ref{conclusion} with possible future directions.


\section{Related Work} \label{relatedwork}

In the literature, the image colorization problem is mainly considered in two categories as follows.

\textbf{Colorization With Guidance.} In this approach, a user interaction or a set of 
guidance pixels are asked to provide a prior information to the colorization system in 
order to obtain realistic results. In \cite{levin2004colorization}, user provided scribbles 
are used to colorize the neighboring pixels that have similar intensity values. This method 
eliminates the need of the object segmentation and considers the problem as an optimization 
procedure that minimizes a quadratic cost function. The method proposed in 
\cite{ironi2005colorization} uses example image segments and transfers the colors from 
segments into the grayscale areas by keeping the spatial coherency high. An interactive 
colorization algorithm, which is proposed in \cite{yatziv2006fast}, assigns weights for 
user scribbles and combines them for final colorization. In \cite{chia2011semantic}, the 
user is asked to provide the localization and labelling for the salient foreground object 
to be colorized and then the object is colorized by using reference images retrieved from 
internet. And \cite{gupta2012image} transfers the color information into the grayscale 
images from semantically similar reference images by using superpixel representation to 
speed up the colorization and to have better spatial consistency. The main drawback of the 
guided colorization algorithms is that they are highly dependent to user feedback or 
reference example images and this causes the algorithms to fail generalizing well for all 
type of colorizations.

\textbf{Colorization With No Guidance.} The methods that apply colorization without any 
feedback fall into this category and they colorize the given grayscale images automatically 
in an end-to-end fashion. In \cite{charpiat2008automatic}, the grayscale pixels are considered 
at local and global levels by estimating the multimodal color distribution and applying 
graph-cut algorithm respectively. The method in \cite{cheng2015deep} proposes a deep 
colorization model combined with adaptive image clustering and joint bilateral filter for 
global consideration and colorization enhancement respectively. In \cite{larsson2016learning}, 
as similar to \cite{charpiat2008automatic,cheng2015deep}, a deep model is trained to 
generate the per-pixel color histogram that represents the multimodal color distribution. 
The method in \cite{iizuka2016let} proposes an end-to-end CNN model that combines the local 
features with global information. In \cite{zhang2016colorful}, as similar to \cite{iizuka2016let}, 
a CNN model is constructed with the class-rebalancing mechanism in order to provide the color 
diversity for the ill-posed nature of the colorization problem. The automatic colorization 
methods provide end-to-end fully colorization mechanisms while they may be lack of solving 
the ambiguity in case of multiple choice of colorization.


\section{Proposed Method} \label{proposedmethod}

In this section, the design of the ColorCapsNet model is explained in detail. First, the 
pre-processing steps such as colorspace selection and data preparation are explained. Then, 
the modifications on topology and optimization procedures are presented. Finally, the 
selection of critical parameters, which affects the colorization performance, are discussed.

\subsection{Data Pre-processing} \label{preprocess}

\textbf{Colorspace.} In this study, as in \cite{charpiat2008automatic,gupta2012image,
larsson2016learning,zhang2016colorful}, CIE Lab colorspace is used to represent the color 
images. CIE Lab is made of three channels as L, a an b. L represents the lightness/luminance 
whereas a and b are the chrominance. CIE Lab is a perceptually linear colorspace as it 
establishes a mapping between the colors in Euclidean space and the colors in human perception. 
Therefore, it's more suitable than other colorspaces for colorization task and preferred 
to represent the colors in this study as well.

\begin{figure}[h!]
  \centering
  \begin{subfigure}[b]{0.80\linewidth}
    \includegraphics[width=\linewidth]{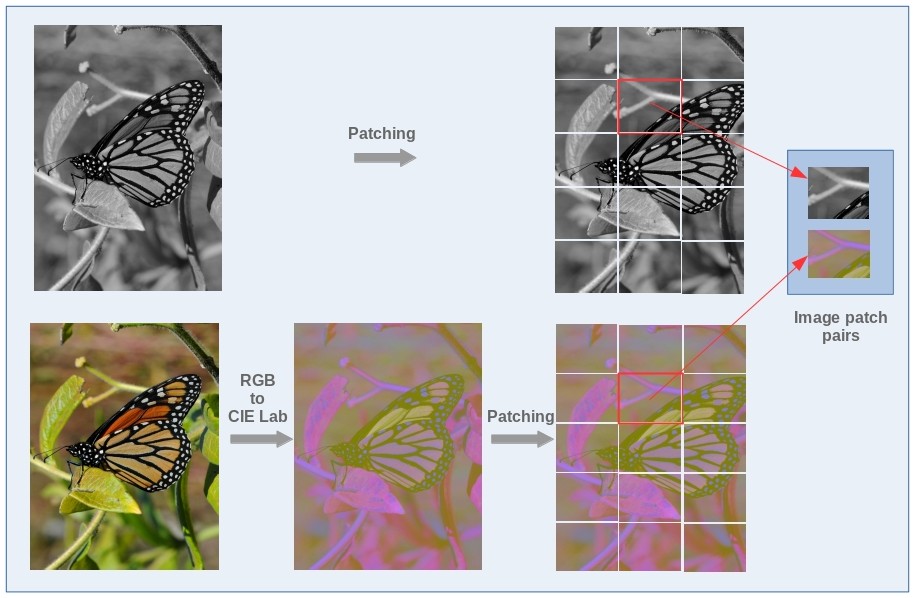}
  \end{subfigure}
  \caption{Data pre-processing flow.}
  \label{fig:data}
\end{figure}

\textbf{Data Representation.} The ColorCapsNet maps a grayscale image patch to the 
corresponding color image patch in CIE Lab colorspace. Therefore, during both training and 
testing phases, the data must be fed into the ColorCapsNet accordingly. For training phase, 
first, each color image given in RGB colorspace is converted into CIE Lab colorspace. Then, 
both the color and corresponding grayscale images are sliced into the $nxn$ square patches, 
where $n$ is a pre-defined value (see Section \ref{parameter}), and stacked as image patch 
pairs. In testing phase, the test image in grayscale is sliced as same in training and 
fed into the trained model for colorization. The result is a stack of the predicted color 
patches in CIE Lab colorspace and the patches are put together to make the resulting image 
complete. Finally, the estimated image is converted from CIE Lab into RGB for visual 
perception. The flow for data pre-processing is illustrated in Figure~\ref{fig:data}.

\subsection{Network Design} \label{design}

The first modification is on the feature detection part of the network. In the original 
CapsNet \cite{sabour2017dynamic}, this part consists of one convolutional layer with 256 
filters. These filters have size of 9x9 with stride of 1 and are activated with Rectified 
Linear Unit (ReLU) to feed the feature representation layer named as primary capsule. In 
proposed method, the feature detector is replaced with the first and the second convolutional 
layers of the VGG-19 model \cite{simonyan2014very} and both layers are identical as they 
have 64 convolution filters with size of 3x3, stride of 1 and ReLU activation. These 
convolutional layers, on contrary to the rest of the network, are initialized by transferring 
the weights of first two layers from the pre-trained VGG-19 model without any freezing 
process. Such an initialization can be regarded as a transfer learning strategy, and, here, 
the purpose is to transfer the low level feature representation capability of the VGG-19 
model into the ColorCapsNet so that it may detect the low level features (corners, edges etc.) 
as good as VGG-19. The performance of detecting low level features has impact on the 
object segmentation that further affects the colorization quality. As seen in 
Figure~\ref{fig:topology}, such modification reduces the train loss from 0.0035 to 0.0033 after 
10 epochs training. In this case, the total number of trainable parameters also reduces by 720000 approximately. 

The second modification is to add Batch Normalization (BN) layer after each convolutional 
layer. The reason of adding BN layer is to reduce the effect of the Internal Covariate 
Shift (ICS) problem stated in the study \cite{ioffe2015batch} that proposes BN as a 
solution to this problem. There are two benefits of applying BN: 1) It speeds up the training 
process, 2) It regularizes the network for better generalization.

BN normalizes the training instance, $x_i$, in the mini-batch 
$B=\{x_1,..,x_b\}$, which is a small representation of train set with the size of 
\textit{b}, so that it has zero mean with unit variance ($\mu=0$,$\sigma^2=1$). In order 
to make the ColorCapsNet ICS-reduced, first, the mean is calculated for the mini-batch 
\textit{B} of the patch pairs as below:

\begin{equation}
\mu_B = \frac{1}{b}\sum_{i=1}^{b}x_i
\end{equation}

\noindent Then, the variance of \textit{B}, $\sigma_B^2$, is calculated:

\begin{equation}
\sigma_B^2 = \frac{1}{b}\sum_{i=1}^{b}(x_i-\mu_B)^2
\end{equation}

\begin{figure*}[h!]
  \centering
  \begin{subfigure}[b]{0.80\linewidth}
    \includegraphics[width=\linewidth]{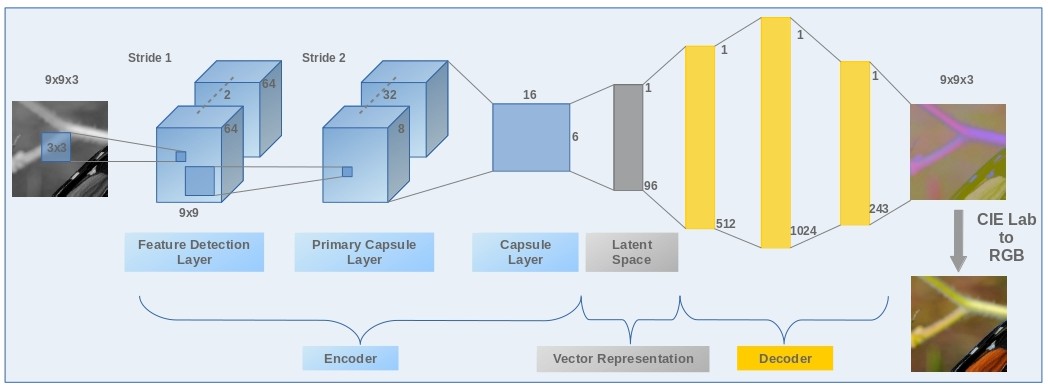}
  \end{subfigure}
  \caption{The ColorCapsNet topology.}
  \label{fig:model}
\end{figure*}

\noindent Finaly, \textit{B} is normalized to zero mean with unit variance to reduce the ICS:

\begin{equation}
\hat{x_i} = \frac{x_i-\mu_B}{\sqrt{\sigma_B^2}}
\end{equation}

BN is applied to the first two convolutional layers of the feature detection part and the 
right after the convolutional layer in the primary capsule part. The effect of applying BN 
can be examined in Figure~\ref{fig:topology}.

The third modification is for the number of capsule in the classification layer 
(capsule layer). In the original CapsNet topology, the number of capsule is selected as 10 
to address the 10-class digit classification task. Here, because the colorization is an image 
generation task rather than classification, the parameter of the capsule number in the capsule layer 
should be adapted to this problem. For this parameter, moving from 10 to 6 doesn't change 
the loss so much as seen in Figure~\ref{fig:topology} but it dramatically reduces the 
number of trainable parameters by 2390000 approximately. Therefore, the number of the capsule 
is selected as 6 in the capsule layer.

The resulting topology is illustrated in Figure~\ref{fig:model}. As seen in the figure, 
the topology resembles an autoencoder that is made of an encoder and decoder network. 
There is a vector representation, which is defined in latent space, between encoder and 
decoder networks and it represents the hidden variables related with colorization task.

\subsection{Optimization Procedure} \label{optimization}

In the original CapsNet model, the margin loss is used to minimize the loss during training. 
The margin loss is defined for each capsule in the classification layer (capsule layer) as 
below:

\begin{equation}
L_c = T_c max(0, 0.9 - \|v_c\|)^2 + \lambda(1 - T_c)max(0,\|v_c\| - 0.1)^2
\end{equation}

\noindent where $\lambda=0.5$, $v_c$ is the output of the capsule \textit{c} and $T_c=1$ when 
the capsule representing the class is activated. The total loss over all class capsules 
are as follows:

\begin{equation}
L = \sum_{c=1}^{C}L_c
\end{equation}

In the ColorCapsNet mode, because the optimization is to minimize the difference between 
real and generated color images, the objective function for loss is defined as Mean Squared 
Error (MSE) as below:

\begin{equation}
MSE = \frac{1}{YX}\sum_{y=1}^{Y}\sum_{x=1}^{X}[I(x,y)-\hat{I}(x,y)]^2
\end{equation}

\noindent where $I(x,y)$ and $\hat{I}(x,y)$ are the corresponding real and generated color 
image pixel values respectively. As stated in \cite{zhang2016colorful}, $L_2$/$MSE$ is the 
proper loss function for CIE Lab colorspace because it defines the chroma in the Euclidean 
space and it's effective to minimize $MSE$ in this space.

As in the CapsNet, the Adam optimizer \cite{kingma2014adam} is leveraged as the optimization 
method during forward-backward pass with the learning rate of 0.001, $\beta_1$ of 0.9 and 
$\beta_2$ of 0.999. 

\subsection{Parameter Selection} \label{parameter}

\begin{figure*}[h!]
  \centering
  \begin{subfigure}[b]{0.30\linewidth}
    \includegraphics[width=\linewidth]{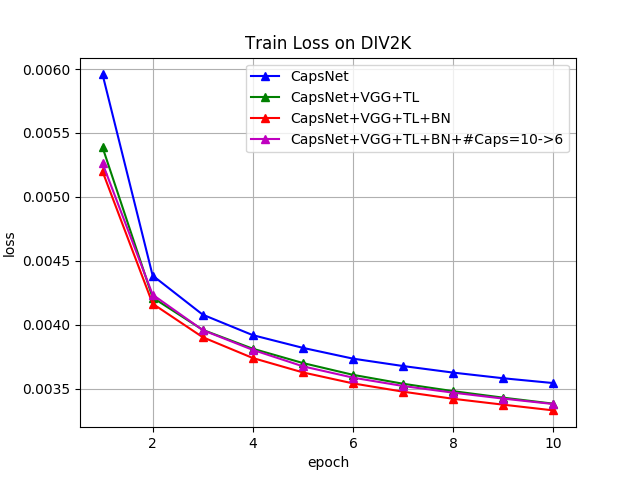}
    \caption{Topology}
    \label{fig:topology}
  \end{subfigure}
  \begin{subfigure}[b]{0.30\linewidth}
    \includegraphics[width=\linewidth]{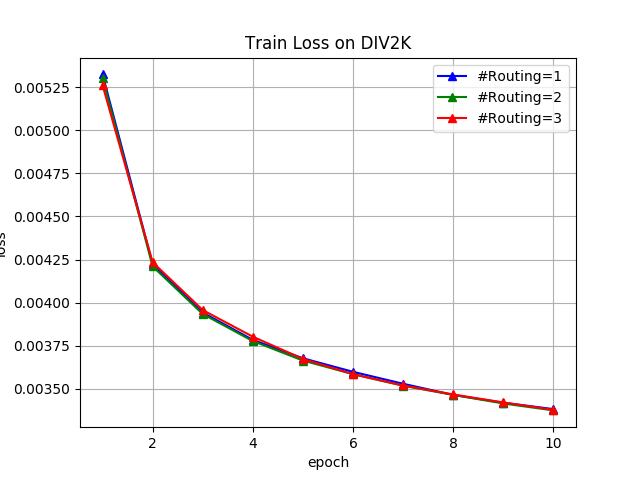}
    \caption{Number of routings}
    \label{fig:routing}
  \end{subfigure}
  \begin{subfigure}[b]{0.30\linewidth}
    \includegraphics[width=\linewidth]{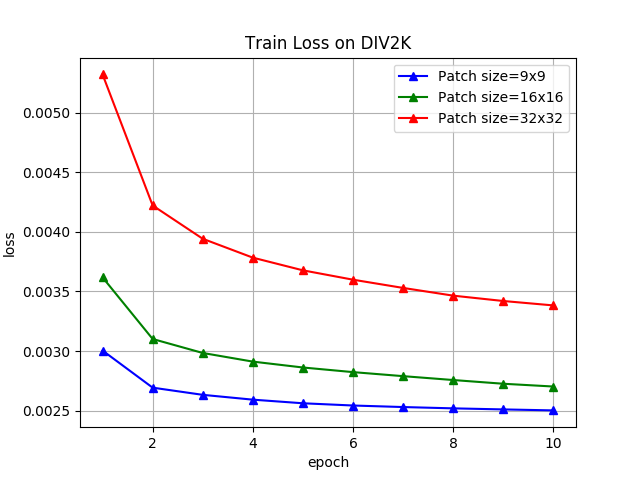}
    \caption{Patch size}
    \label{fig:patch}
  \end{subfigure}
  \caption{Comparative analysis for network design of the ColorCapsNet.}
  \label{fig:comparative_design}
\end{figure*}

\textbf{Number of Routings.} The routing is one of the most critical hyperparameters in the 
CapsNet topology. The routing is an iterative procedure for information transfer between the 
capsules in different layers. In the routing mechanism, the capsule in the lower layer is 
connected to the activated capsule in the upper layer for transferring its output. This mechanism 
is called as \textit{"routing-by-agreement"}. According to \cite{sabour2017dynamic}, the 
\textit{"routing-by-agreement"} leverages the information about the shape of the object in 
pixel level so that the segmentation could be achieved while moving from locality to global 
extent. The ColorCapsNet also exploits the idea of routing in order to segment objects internally 
and to colorize them. The number of the iterations during routing agreement should be 
selected carefully in order to achieve proper information transfer between capsules in 
different layers. In \cite{sabour2017dynamic}, the best performance is obtained by 
routing 3 times. In Figure~\ref{fig:routing}, it can be observed that the change in the 
number of routing doesn't effect the colorization performance. The drawback of 
using iterative routing is that it increases the time complexity of training as the number 
of routing goes up. By considering this fact, the number of routings is selected as 1 in 
order to accelerate the training process.

\textbf{Patch Size.} The size of patch is another critical parameter for successful 
colorization because it defines the boundary of the local structures and leads to better 
understanding of exposing them. In this study, three different patch sizes are considered 
for colorization task: 9x9, 16x16 and 32x32. 9x9 and 32x32 are selected as the minimum and 
the maximum patch sizes respectively because 9x9 is the theoretical lower bound as 
input and the patch sizes bigger than 32x32 cause visual discontinuities at border of 
adjacent patches. In Figure~\ref{fig:patch}, it's obviously seen that the train loss 
decreases with the patch size goes down exponentially. Therefore, the patch size is 
selected as 9x9 for this problem.


\section{Experimental Analysis} \label{experimentalanalysis}

In this section, the performance of the ColorCapsNet is quantitatively evaluated by testing 
with validation and test sets from DIV2K dataset. The results are reported with well-known 
evaluation metrics, PSNR and SSIM. Some perceptual results are also shown from validation 
and test sets to demonstrate the effectiveness of the ColorCapsNet on the fully automatic 
colorization task.

\subsection{Datasets}

ILSVRC 2012 \cite{deng2009imagenet} and DIV2K \cite{agustsson2017ntire} are two datasets 
used in this study to model the colorization. The reason of using ILSVRC 2012 is to 
supervise the ColorCapsNet with general color distribution of the objects in the dataset. 
This helps the network to have prior information about object colors before trained on DIV2K. 
Once, the ColorCapsNet is trained on ILSVRC 2012, as next step, DIV2K dataset is used for 
training to get the final colorization model. Both datasets are pre-processed to have 
colorized-grayscaled image pairs in pre-defined patch size.

The validation and test sets of the ILSVRC 2012 dataset are used in this study as follows: 
First, both sets are merged with ending up 150000 RGB images. Then, these 150000 images are 
converted into the images in grayscale and CIE Lab. Finally, the corresponding grayscale 
and CIE Lab images are divided into the 9x9x3 patches to train the network. The total number 
of patches is 26536446.

DIV2K dataset has 794 grayscale and 800 RGB images in the train set, and it has 100 
grayscale/RGB pairs in the validation set. First, 794 out of 800 images in the train set 
are used for consistency and there is only RGB to CIE Lab conversion because corresponding 
grayscale images are provided. Grayscale/CIE Lab image pairs are again divided into 9x9x3 
patches with the total number of 6788644 image pairs. The pre-trained network on ILSVRC 
2012 is trained on this dataset for estimating the images from validation set (The 
estimations of the validation images in Figure~\ref{fig:result} are based on this model). Then, 
100 grayscale/RGB pairs from validation set are considered in same manner with ending up 
869196 image pairs as 9x9x3 patches. The pre-trained network on train set of DIV2K is finally 
trained on this validation set to colorize the test images (The test images in 
Figure~\ref{fig:result} are colorized with this model).

\subsection{Evaluation Metrics}

In image colorization domain, the Peak Signal to Noise Ratio (PSNR) and the Structural 
Similarity Index Measure (SSIM) are two widely used evaluation metrics to show the 
effectiveness of the colorization operation. In this study, the colorization performance 
is also evaluated with these metrics.

PSNR is the ratio between the power of the peak signal and the power of the noisy signal 
in terms of the logarithmic scale. It's formulated as below:

\begin{equation}
PSNR = 10\log_{10}(\frac{peak}{MSE})
\end{equation}

\noindent where $peak$ is the power of the peak signal and $MSE$ is the mean squared error 
between the original and the noise signals. The bigger PSNR values mean to the better quality 
and the lower noise in the image. Similarly, in the domain of the image colorization, the bigger 
PSNR indicates the better colorization performance.

SSIM is another image quality metric and used to measure the similarity between real 
and estimated images based on luminance, contrast and structure. It's formulated as below:

\begin{equation}
SSIM(I,\hat{I}) = \frac{(2\mu_I\mu_{\hat{I}}+C_1)(2\sigma_{I\hat{I}}+C_2)}{(\mu_I^2+\mu_{\hat{I}}^2+C_1)(\sigma_I^2+\sigma_{\hat{I}}^2+C_2)}
\end{equation}

\begin{figure*}[h!]
  \centering
  \begin{subfigure}[b]{0.30\linewidth}
    \includegraphics[width=\linewidth]{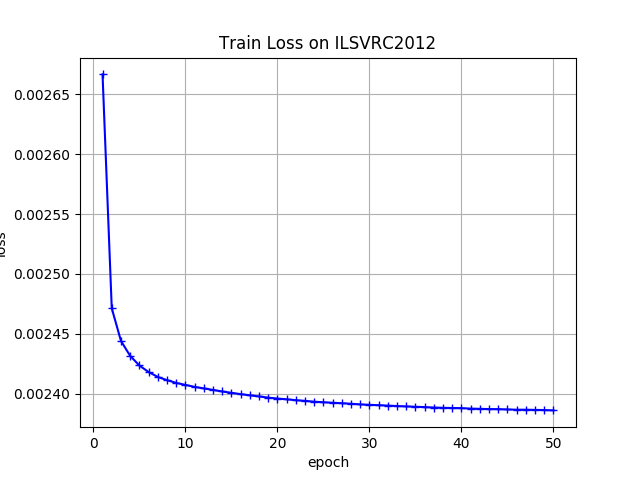}
    \caption{Train loss on ILSVRC 2012}
    \label{fig:ilsvrc}
  \end{subfigure}
  \begin{subfigure}[b]{0.30\linewidth}
    \includegraphics[width=\linewidth]{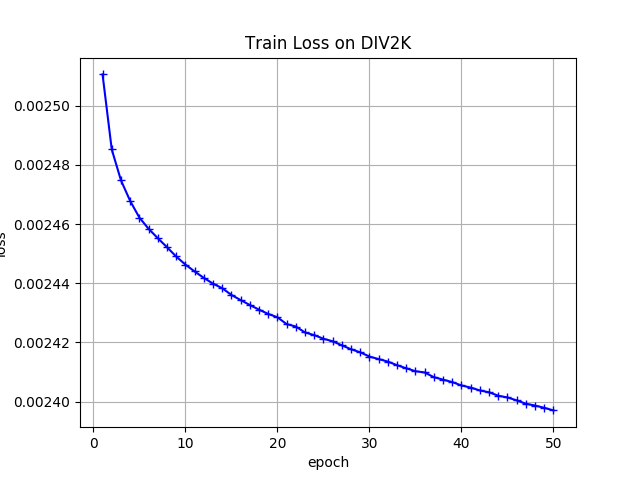}
    \caption{Train loss on DIV2K train set}
    \label{fig:div2ktrain}
  \end{subfigure}
  \begin{subfigure}[b]{0.30\linewidth}
    \includegraphics[width=\linewidth]{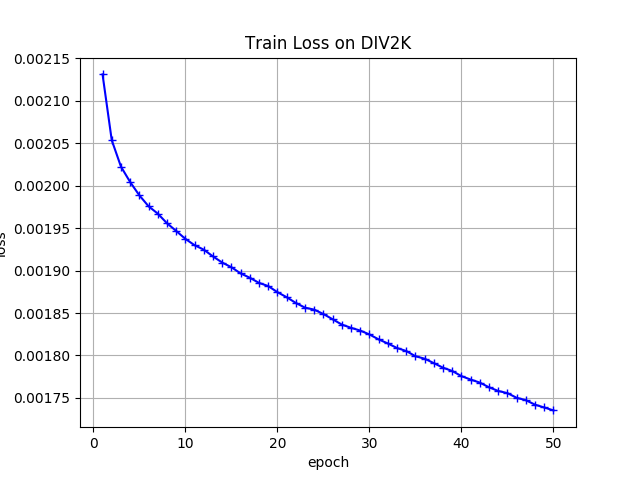}
    \caption{Train loss on DIV2K validation set}
    \label{fig:div2kval}
  \end{subfigure}
  \caption{The train loss performance of the ColorCapsNet.}
  \label{fig:loss}
\end{figure*}

\noindent where $\mu$ is the mean, $\sigma$ is the standard deviation, $\sigma_{I\hat{I}}$ 
is the covariance and $C_1$ and $C_2$ are the contrast related parameters. The bigger SSIM 
indicates the better reconstruction of the real image by the estimated one. For the image 
colorization, the bigger SSIM value means to the better colorization result.

\subsection{Results}

The loss performance during training on ILSVRC 2012, DIV2K train and validation sets is 
shown in Figure~\ref{fig:loss}. Although the training on ILSVRC 2012 is 
saturated in 50 epochs (see Figure~\ref{fig:ilsvrc}), there is still chance to decrease the 
loss after 50 epochs for the training on DIV2K dataset (Figure~\ref{fig:div2ktrain},
Figure~\ref{fig:div2kval}).

For the validation and test phases of the NTIRE 2019 Colorization Challenge \cite{Gu_2019_CVPR_Workshops}, 
the comparative performance of the ColorCapsNet with other methods can be examined in Table~\ref{table:allresult}. 
Proposed method, the ColorCapsNet, has comparable colorization performance on both phases although 
it's just a shallow CapsNet architecture and doesn't contain any complex mechanism for some 
important problems such as spatial coherency. The validation performance for the ColorCapsNet trained 
in 50 epochs is as PSNR of 22.20 and SSIM of 0.88. The test performance in 10 epochs is as PSNR of 21.08 
and SSIM of 0.85 (The testing in 50 epochs couldn't be evaluated because of the unavailability of the colorized 
test data). The worst and the best performances are listed in Table~\ref{table:allresult} as well independently for 
PSNR and SSIM values. In other word, these values are independently the worst and the best values that may or may 
not belong to the same participant.

As seen from Figure~\ref{fig:div2ktrain} and Figure~\ref{fig:div2kval}, it's also possible to continue 
training on DIV2K for better fitting because there is no elbow in the plots after 50 epochs so that it 
has still capacity to converge further. The original CapsNet model \cite{sabour2017dynamic} is trained 
in 1250 epochs with the sign of the convergence at 500 epochs for the digit classification task.


\begin{table}
\begin{center}
\begin{tabular}{|l|c|c|c|}
\hline
Method & Metric & Validation & Test \\
\hline\hline
Worst &  & 16.39/0.55 & 17.96/0.84 \\
Best & \specialcell{PSNR/ \\ SSIM} & 22.73/0.93 & 22.19/0.94 \\
\textbf{ColorCapsNet} &  & $22.20/0.88^1$ & $21.08/0.85^2$\\
\hline
\end{tabular}
\end{center}
\caption{Validation and test results in the NTIRE 2019 Colorization Challenge \cite{Gu_2019_CVPR_Workshops} ($^1:\textit{50}\:epochs$,\,$^2:\textit{10}\:epochs$).}
\label{table:allresult}
\end{table}

\begin{figure*}[h!]
  \centering
  \begin{subfigure}[b]{0.20\linewidth}
    \includegraphics[width=\linewidth]{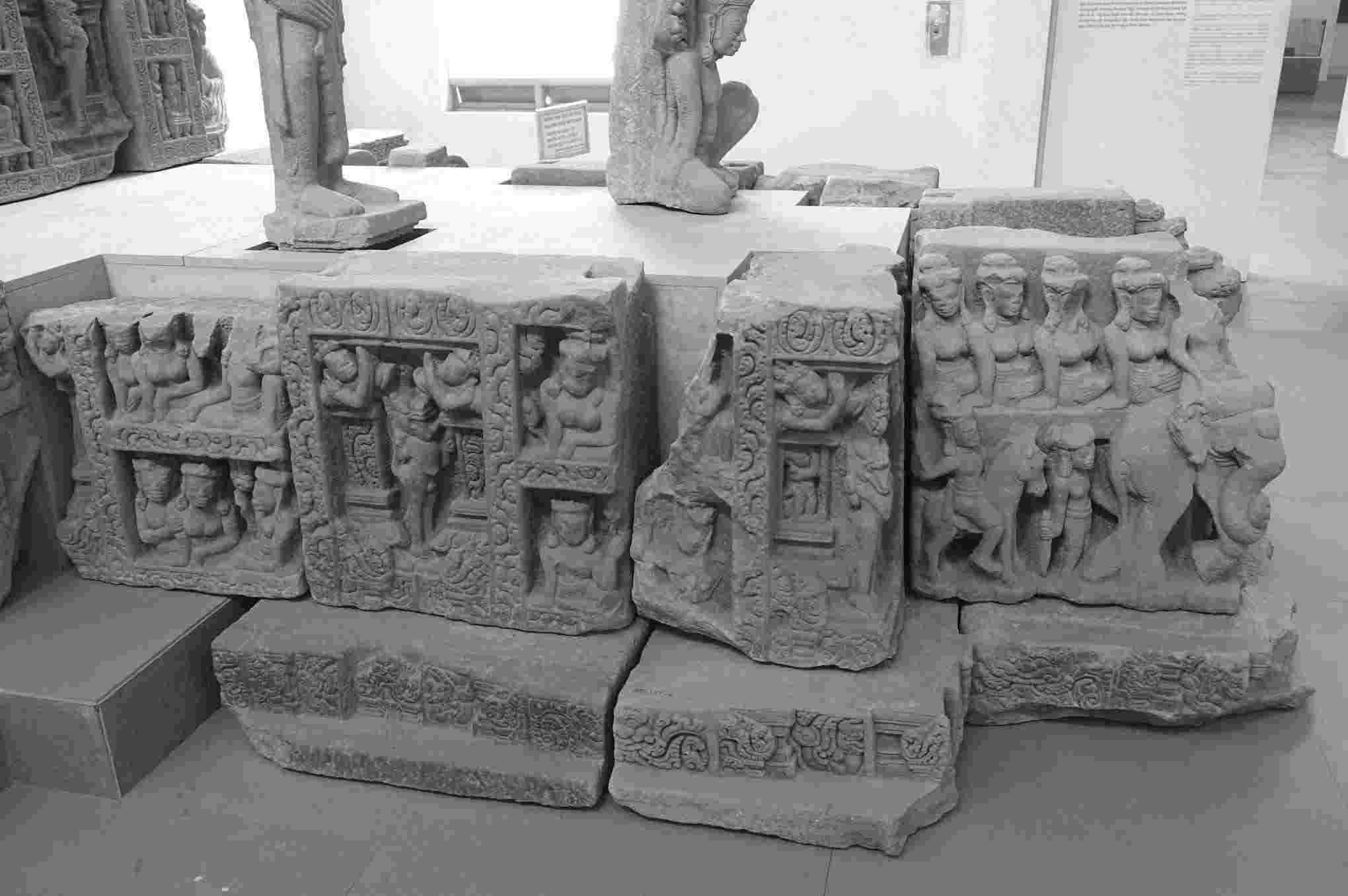}
    \includegraphics[width=\linewidth]{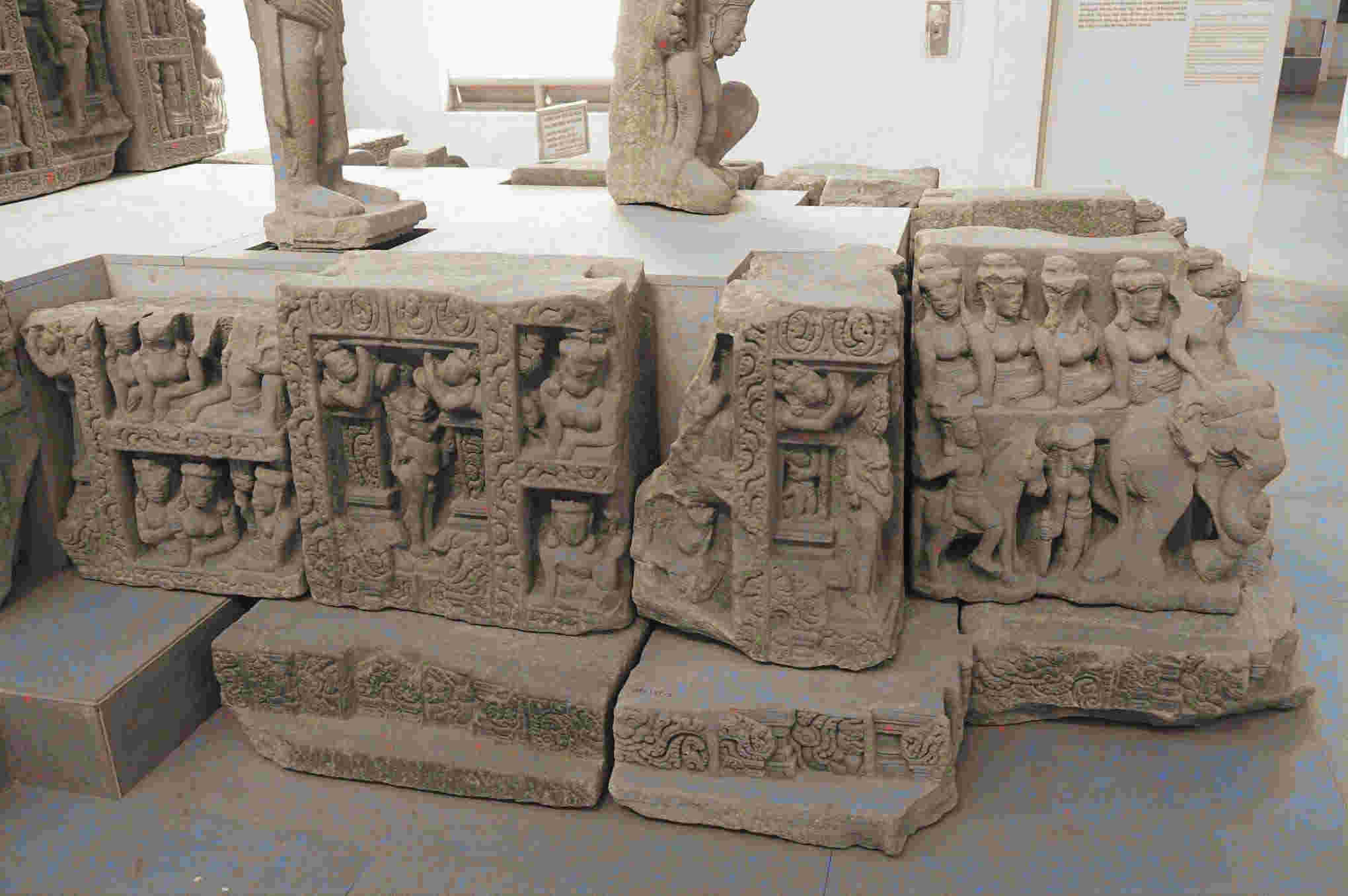}
    \caption{psnr=28.46,ssim=0.92}
  \end{subfigure}
  \begin{subfigure}[b]{0.20\linewidth}
    \includegraphics[width=\linewidth]{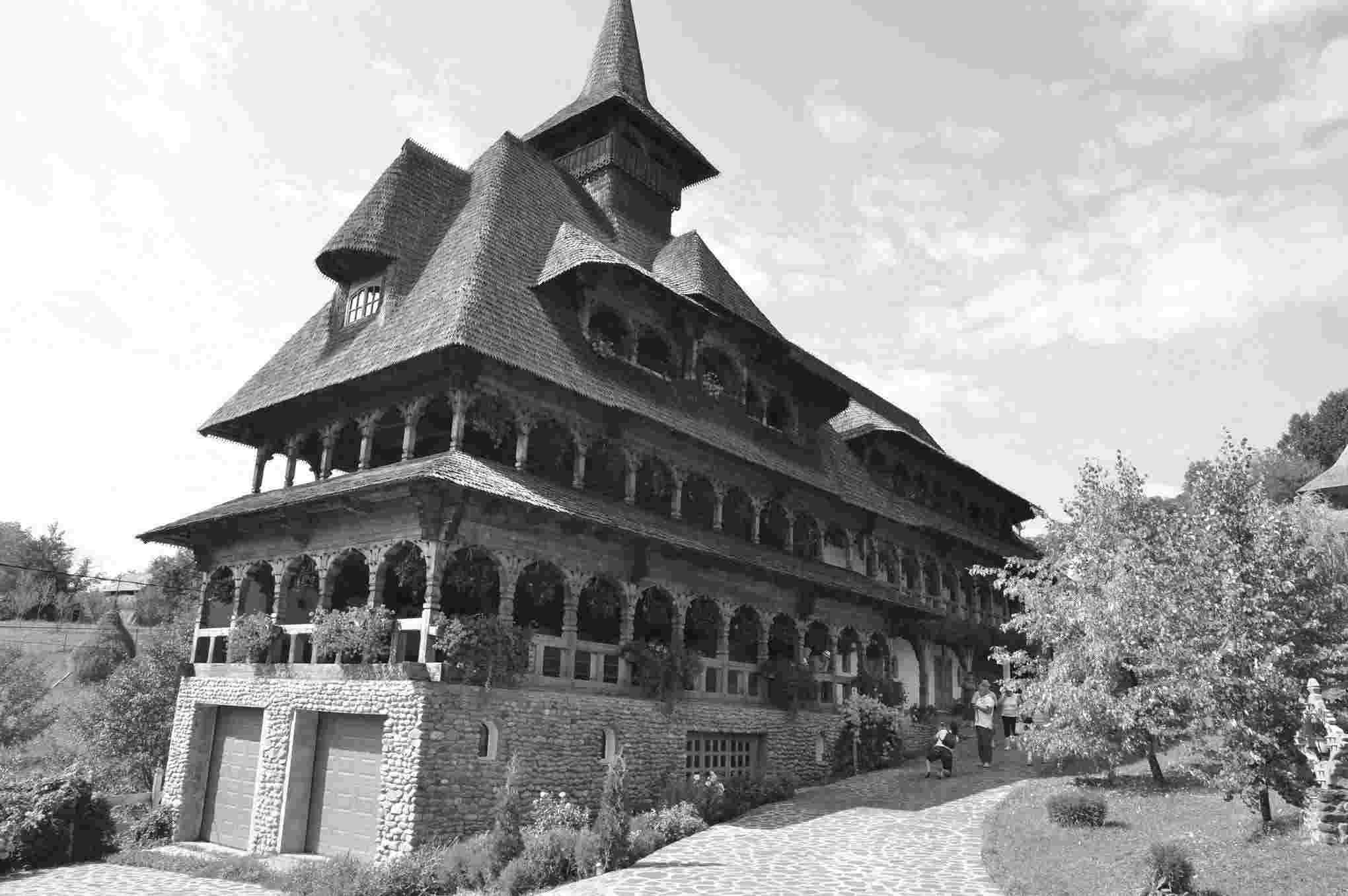}
    \includegraphics[width=\linewidth]{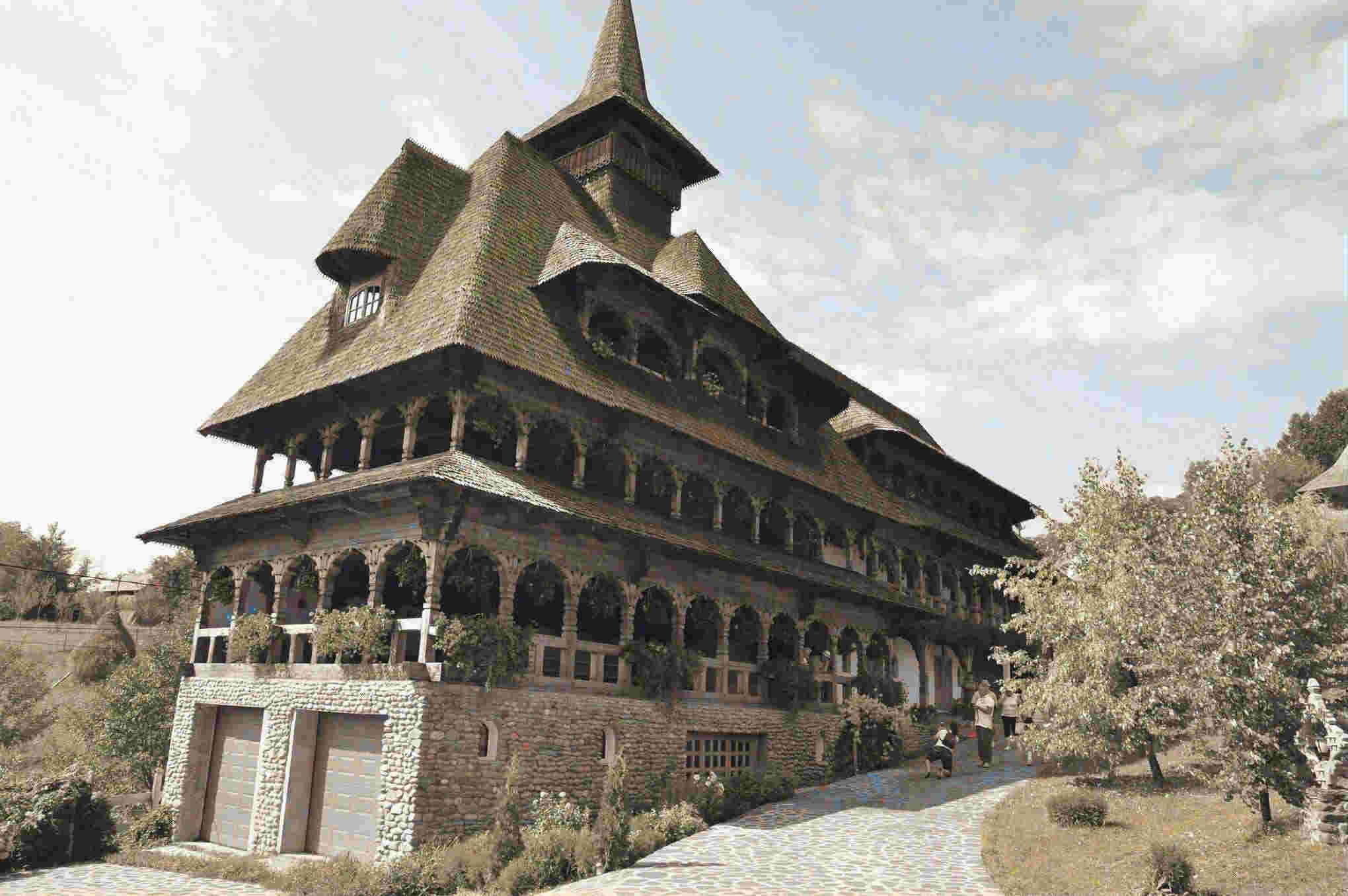}
    \caption{psnr=23.26,ssim=0.94}
  \end{subfigure}
  \begin{subfigure}[b]{0.20\linewidth}
    \includegraphics[width=\linewidth]{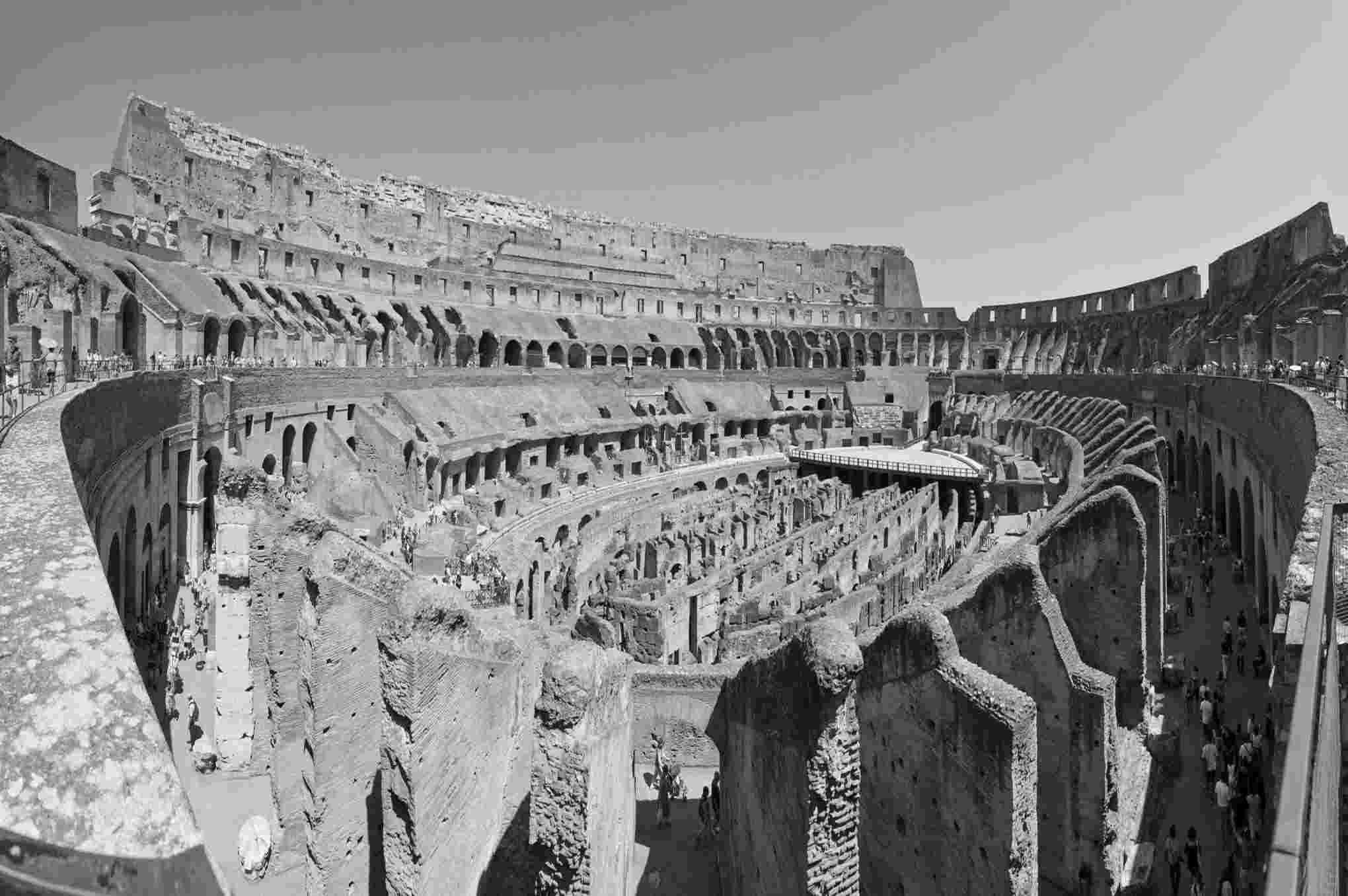}
    \includegraphics[width=\linewidth]{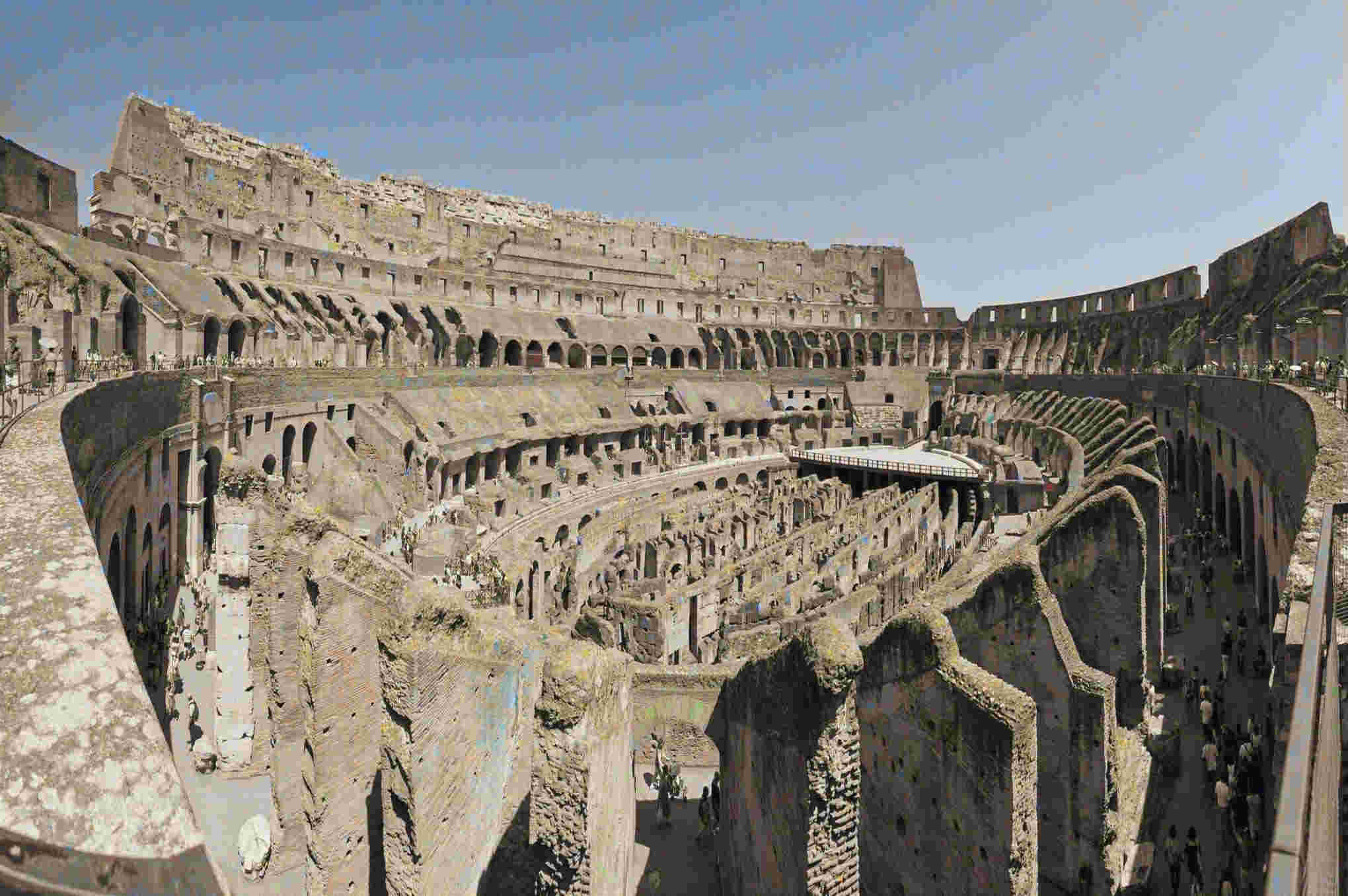}
    \caption{psnr=26.58,ssim=0.94}
  \end{subfigure}
  \begin{subfigure}[b]{0.20\linewidth}
   \includegraphics[width=\linewidth]{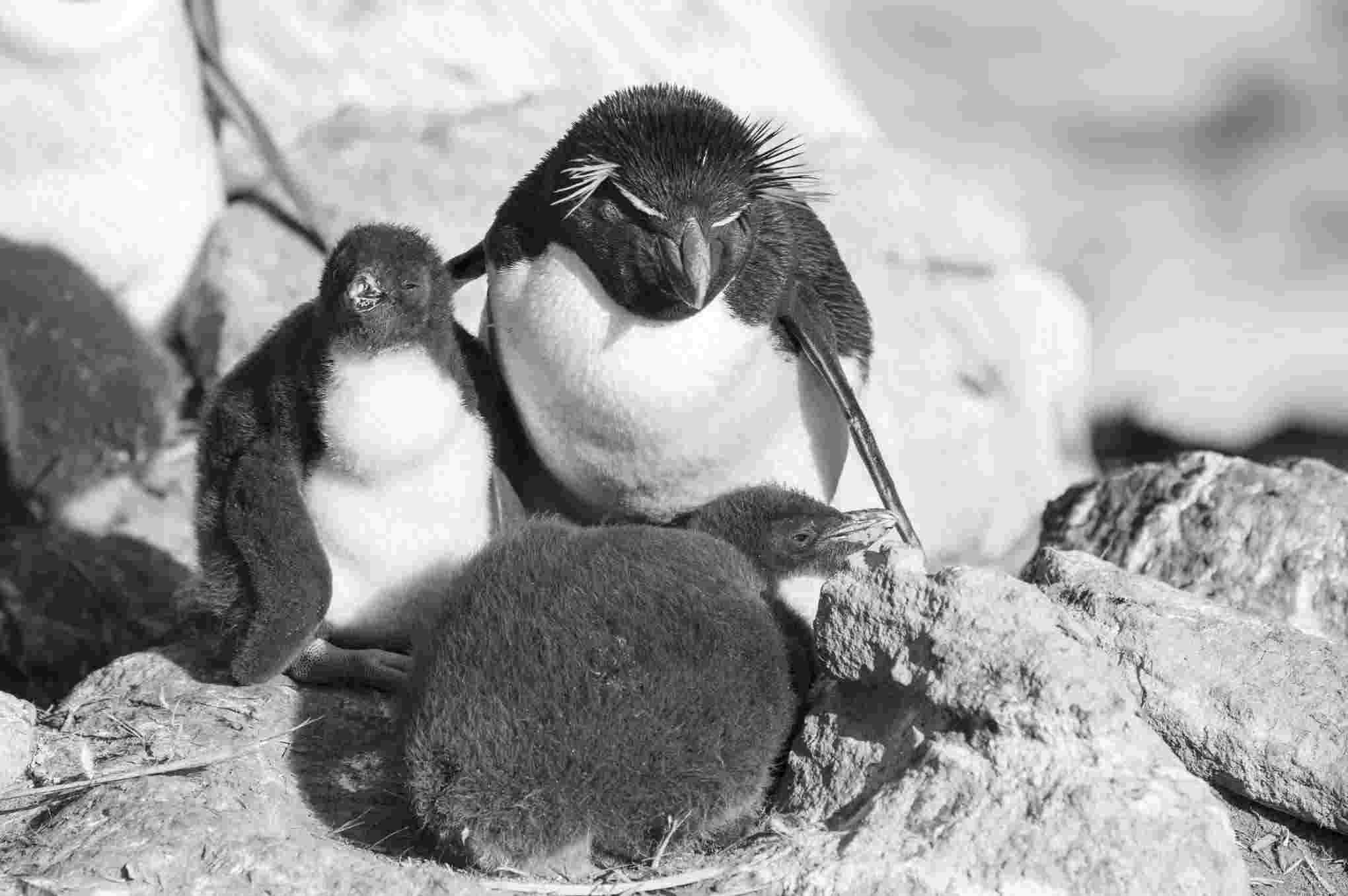}
    \includegraphics[width=\linewidth]{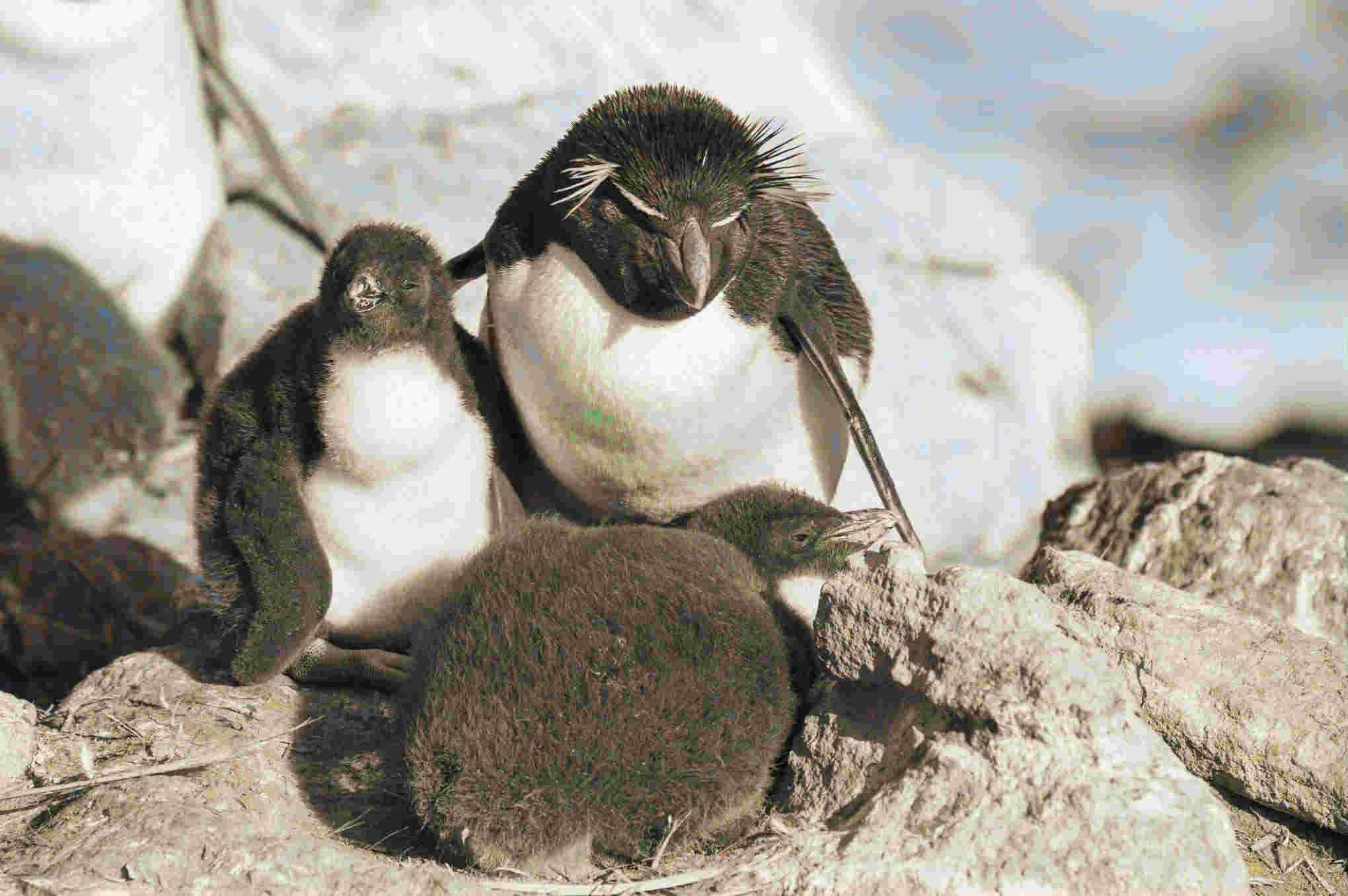}
    \caption{psnr=26.57,ssim=0.95}
  \end{subfigure}
  \begin{subfigure}[b]{0.20\linewidth}
    \includegraphics[width=\linewidth]{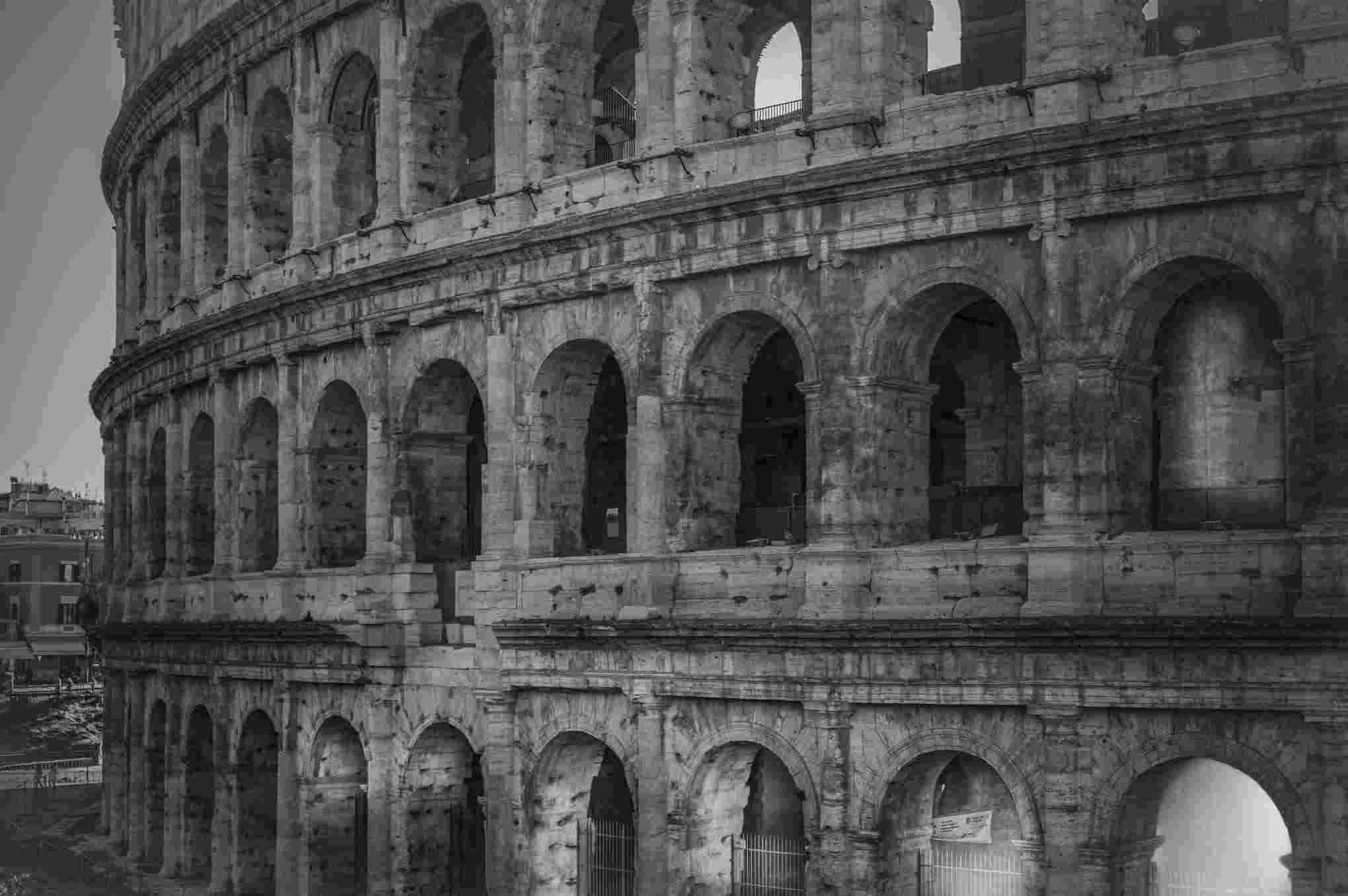}
    \includegraphics[width=\linewidth]{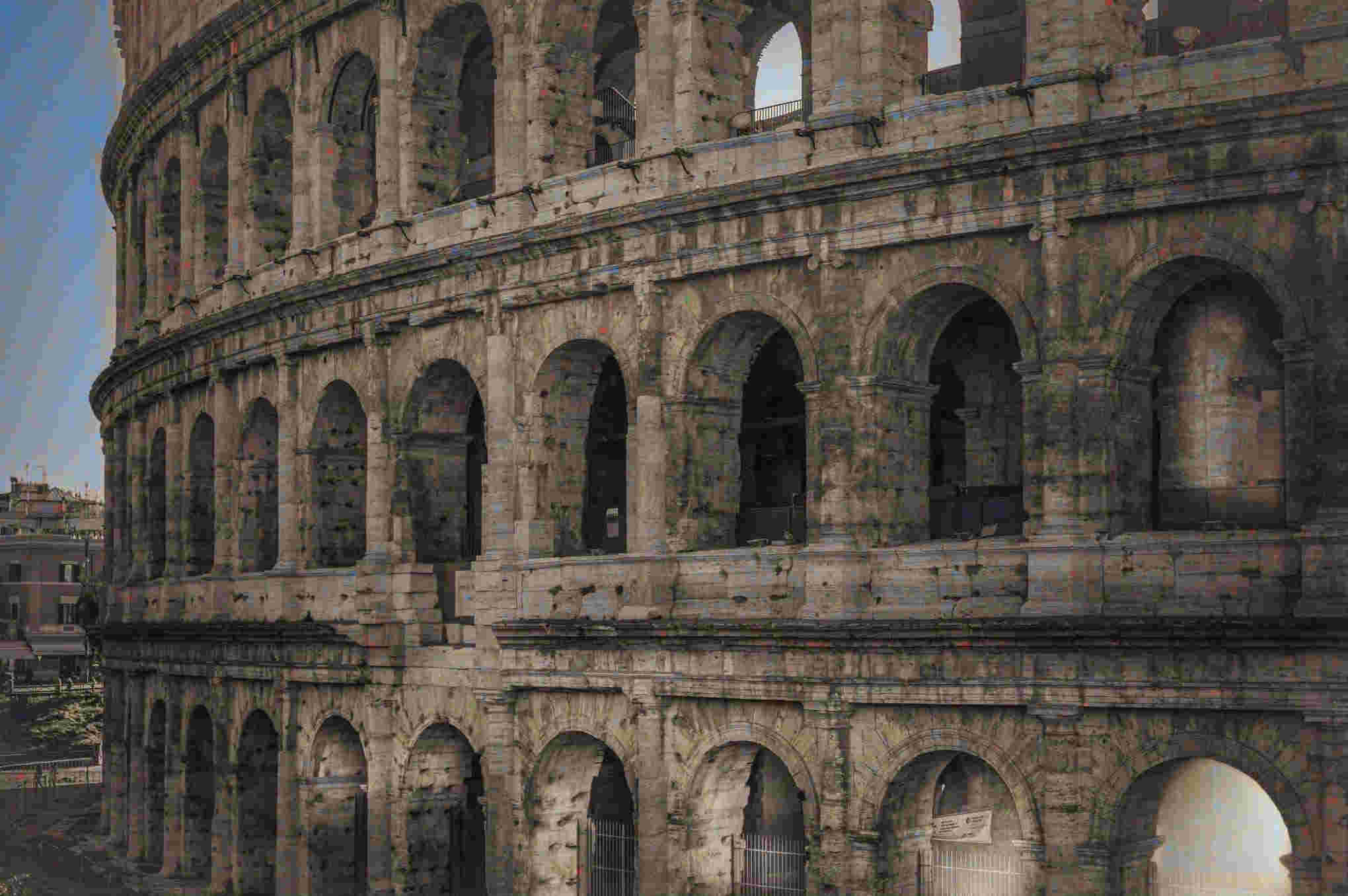}
    \caption{psnr=26.40,ssim=0.95}
  \end{subfigure}
  \begin{subfigure}[b]{0.20\linewidth}
    \includegraphics[width=\linewidth]{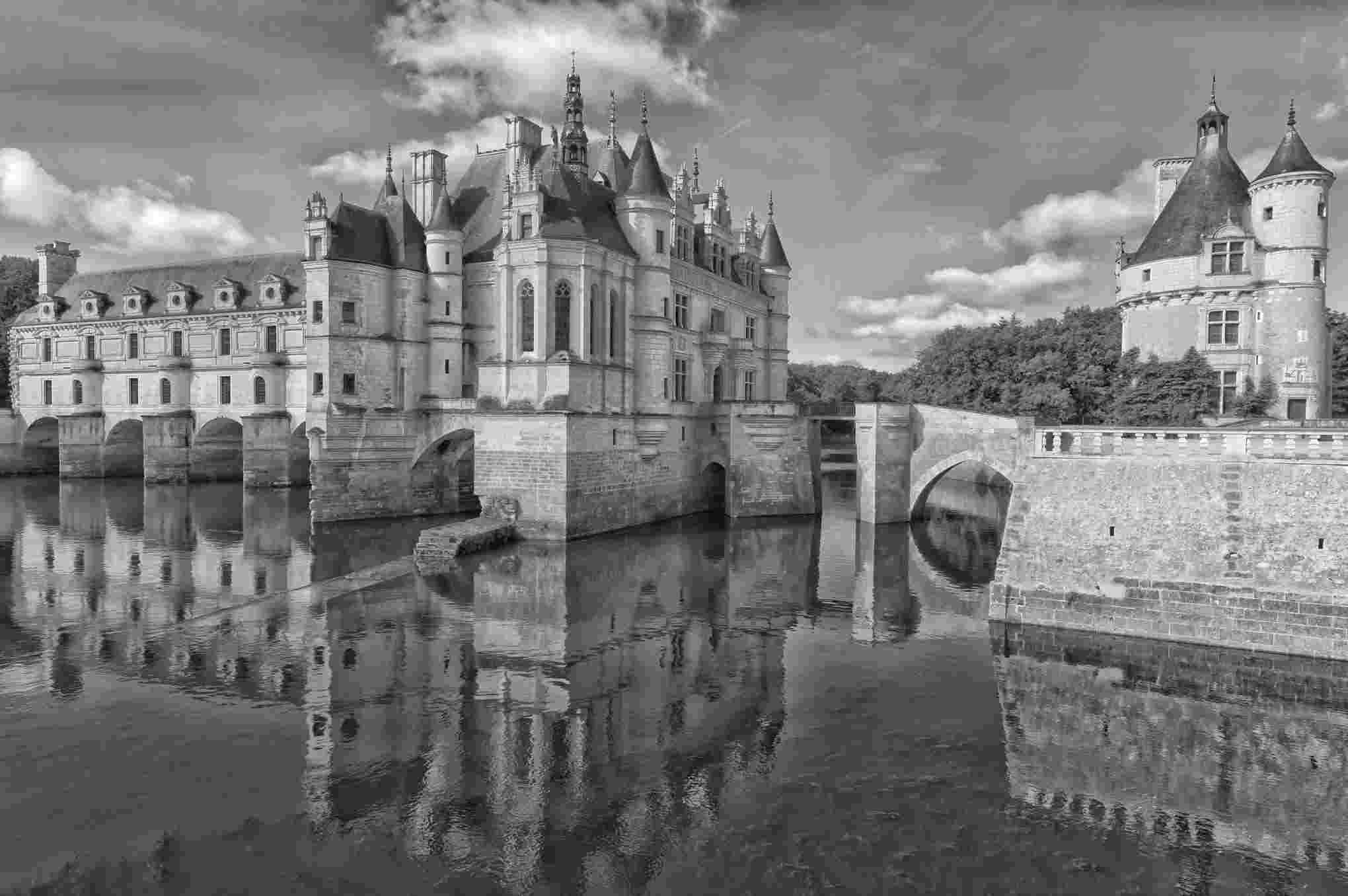}
    \includegraphics[width=\linewidth]{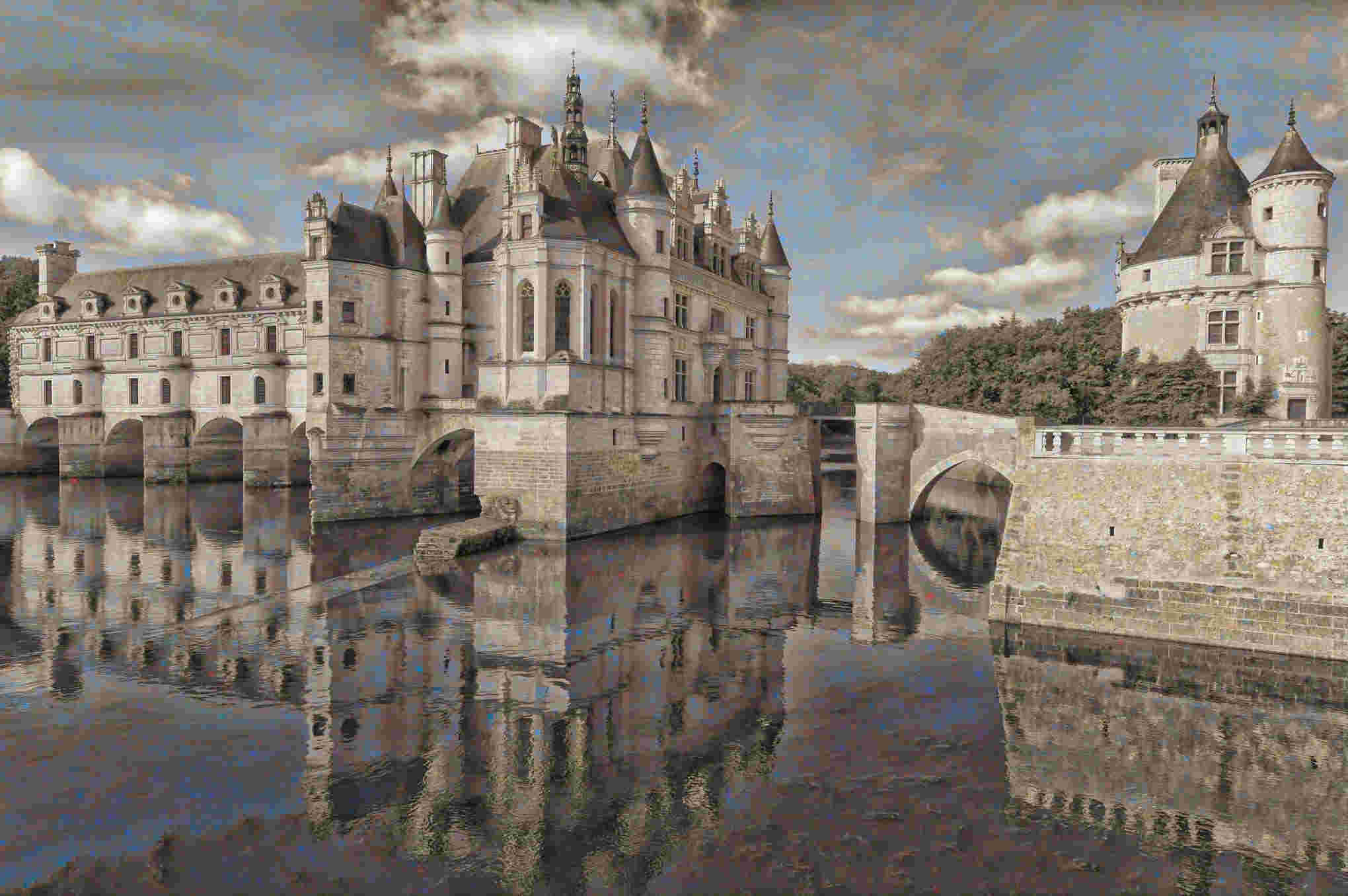}
    \caption{psnr=21.98,ssim=0.89}
    \label{fig:bad1}
  \end{subfigure}
  \begin{subfigure}[b]{0.20\linewidth}
    \includegraphics[width=\linewidth]{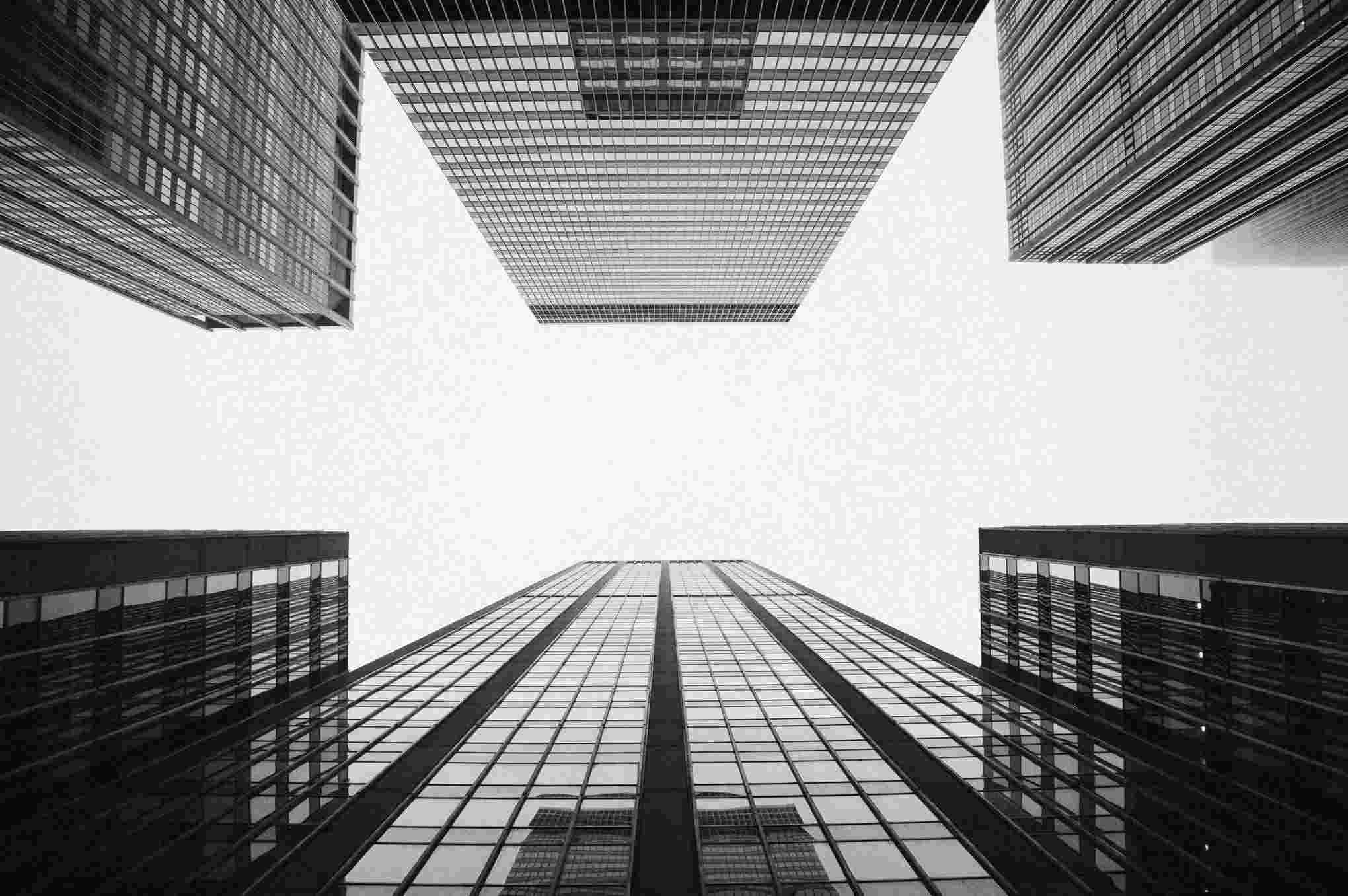}
    \includegraphics[width=\linewidth]{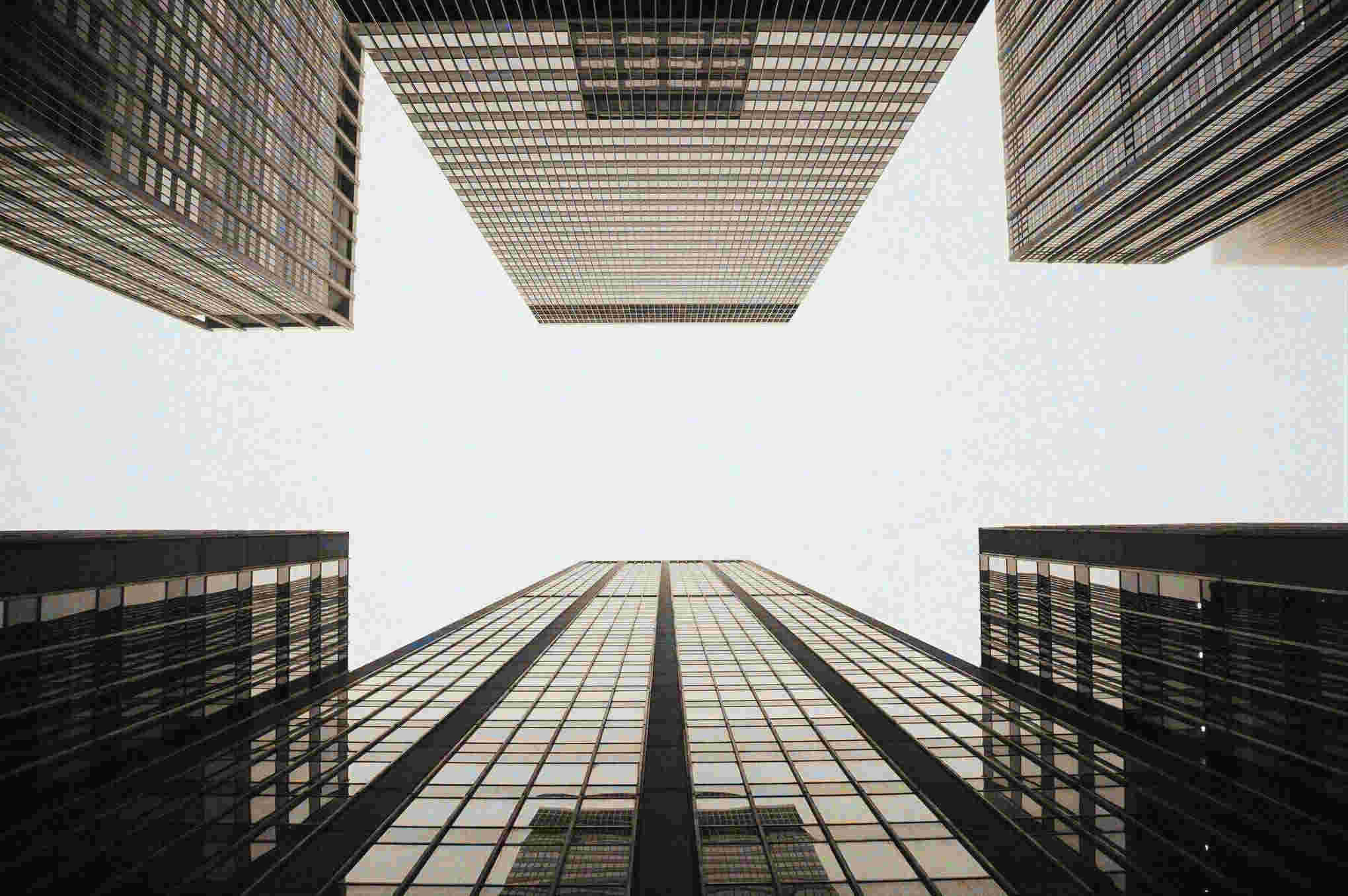}
    \caption{psnr=25.73,ssim=0.93}
  \end{subfigure}
  \begin{subfigure}[b]{0.20\linewidth}
    \includegraphics[width=\linewidth]{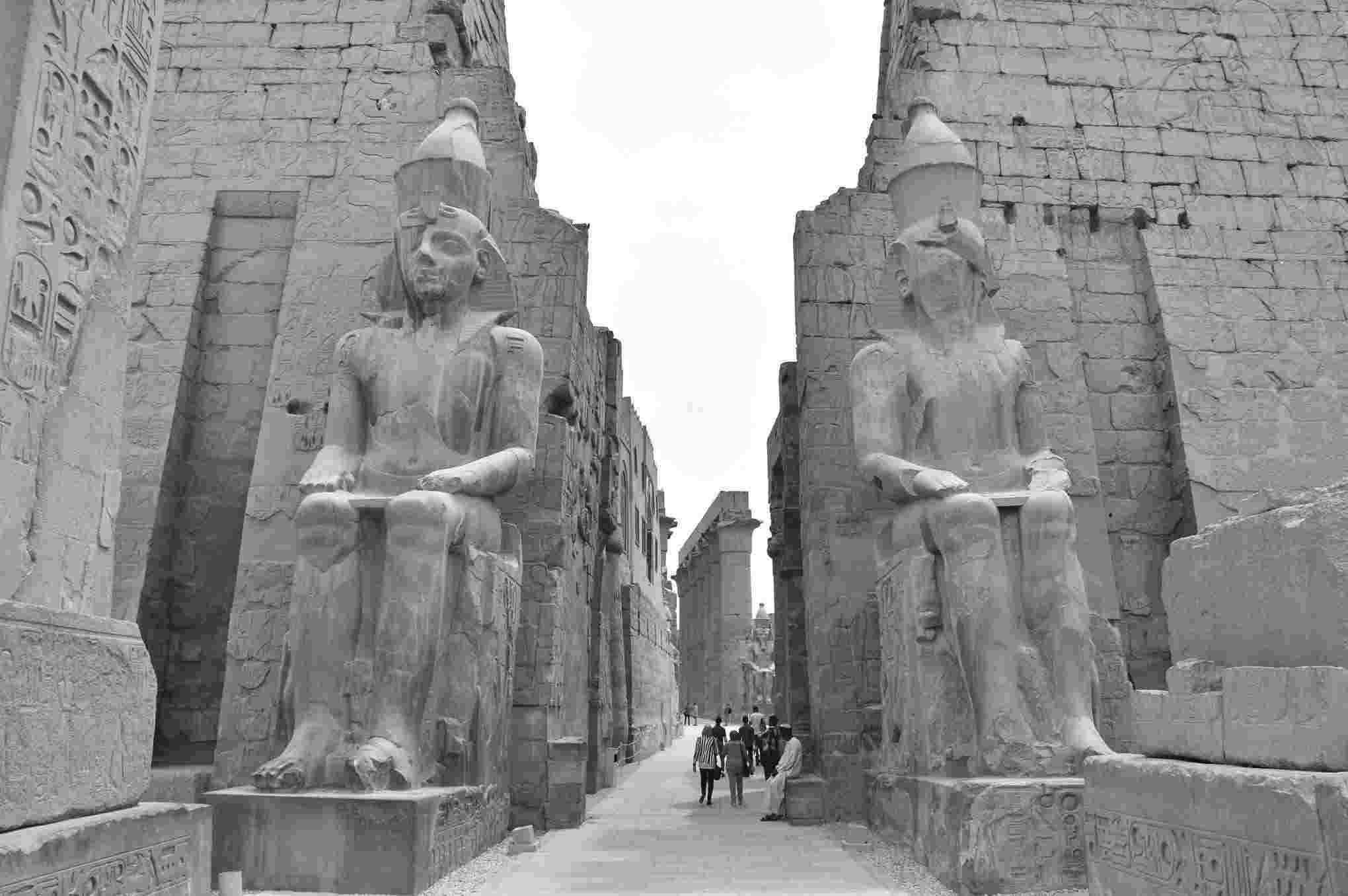}
    \includegraphics[width=\linewidth]{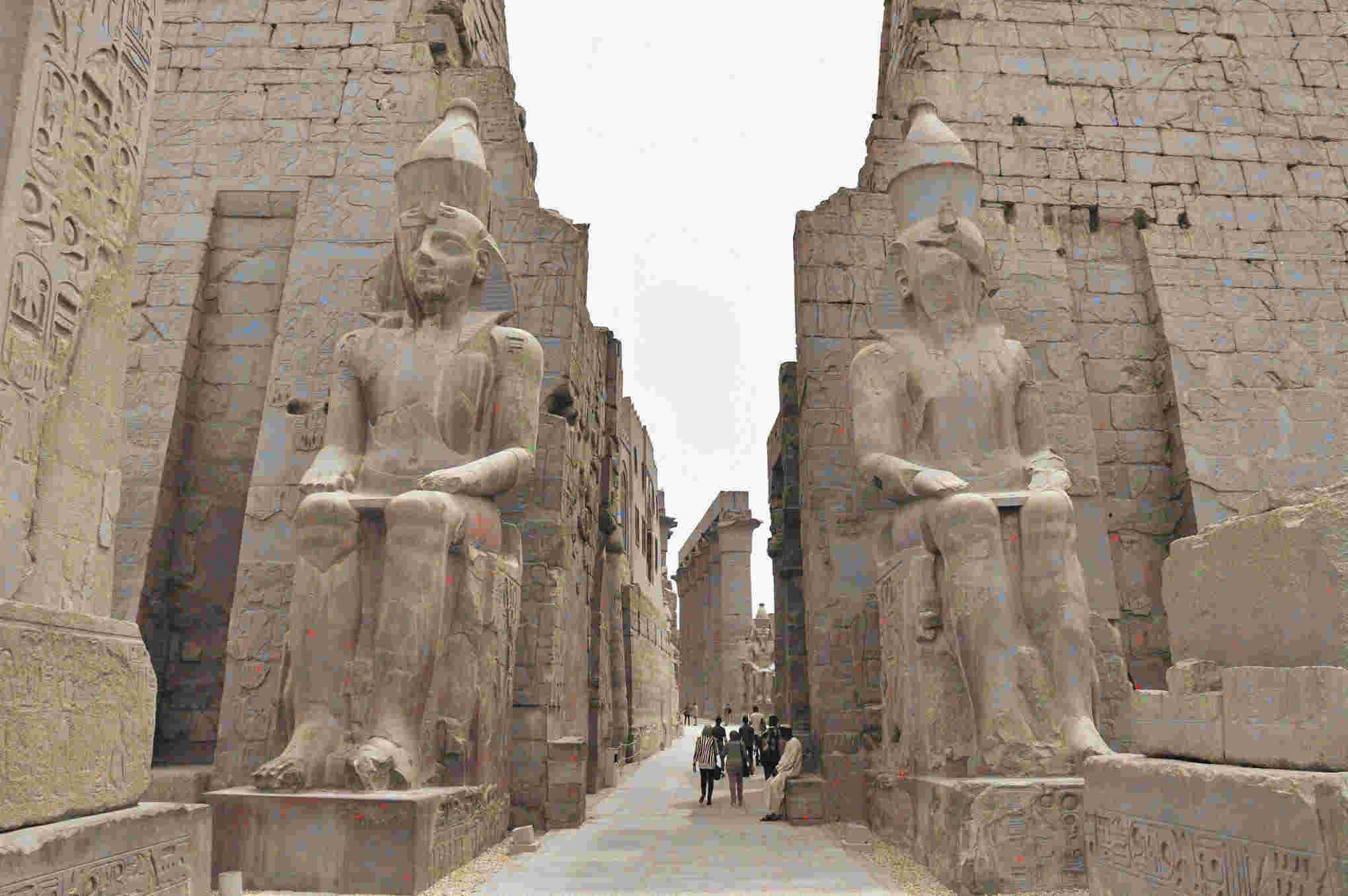}
    \caption{psnr=28.26,ssim=0.94}
  \end{subfigure}
  \begin{subfigure}[b]{0.20\linewidth}
    \includegraphics[width=\linewidth]{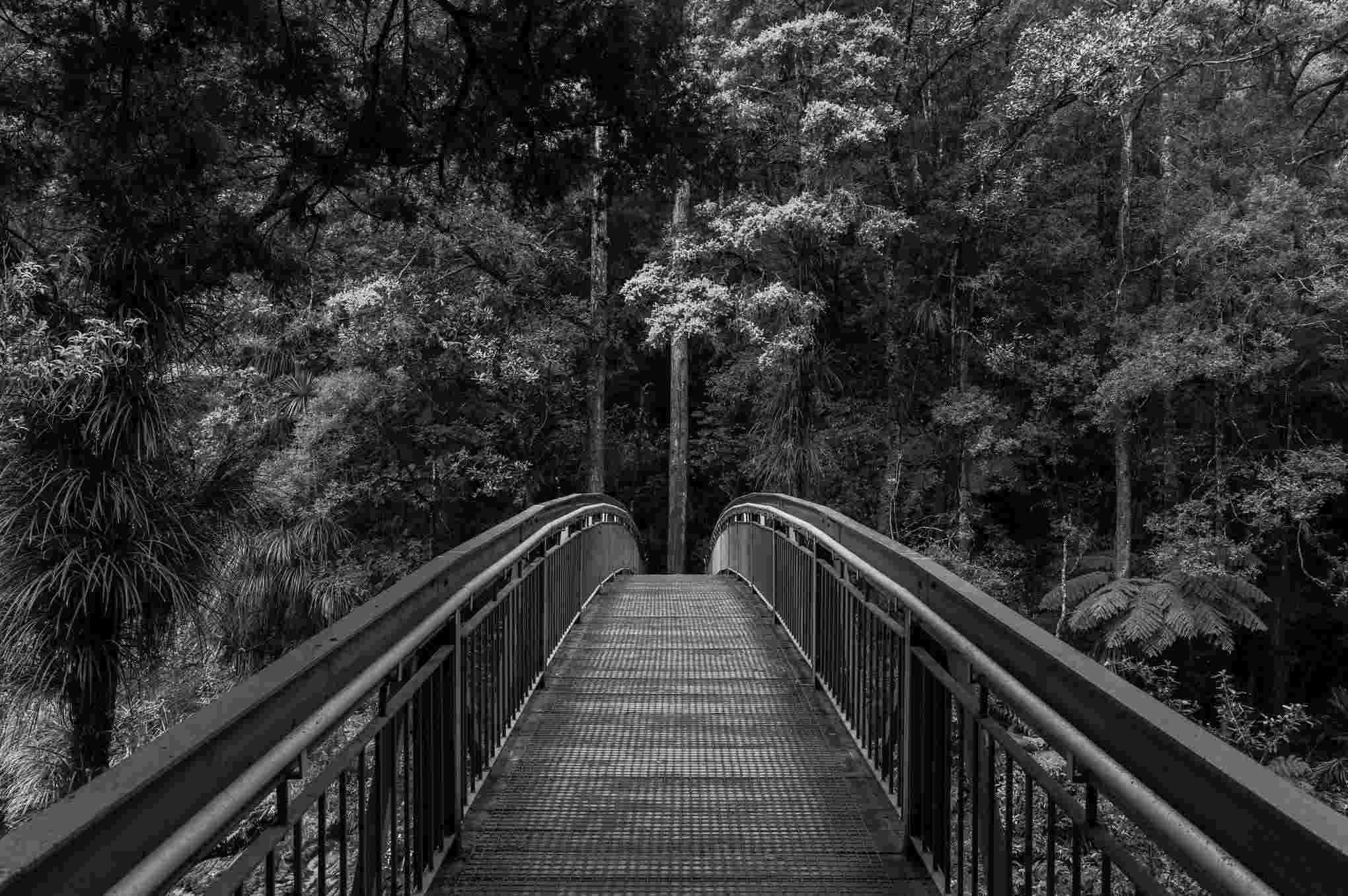}
    \includegraphics[width=\linewidth]{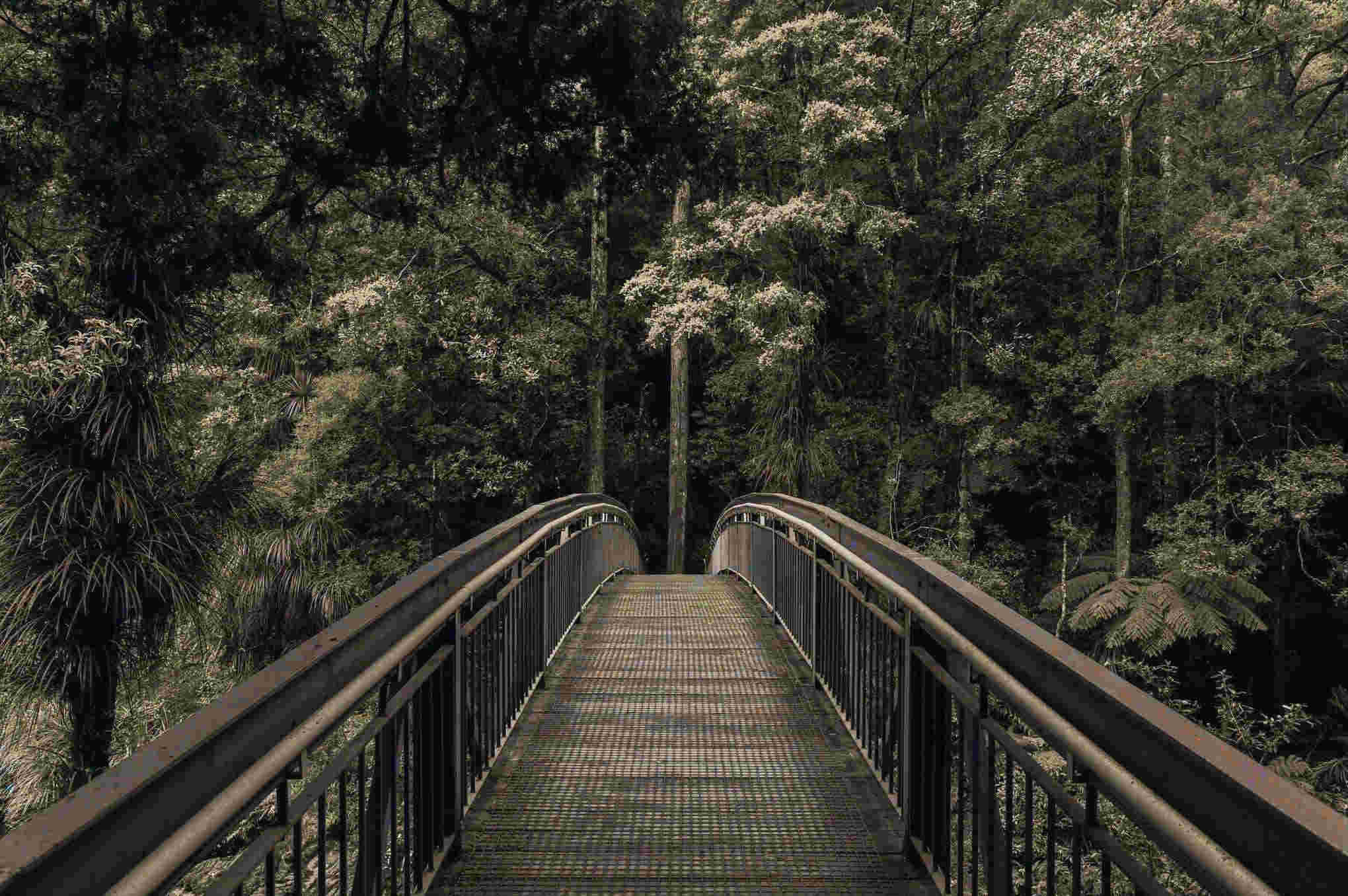}
    \caption{psnr=26.92,ssim=0.91}
  \end{subfigure}
  \begin{subfigure}[b]{0.20\linewidth}
    \includegraphics[width=\linewidth]{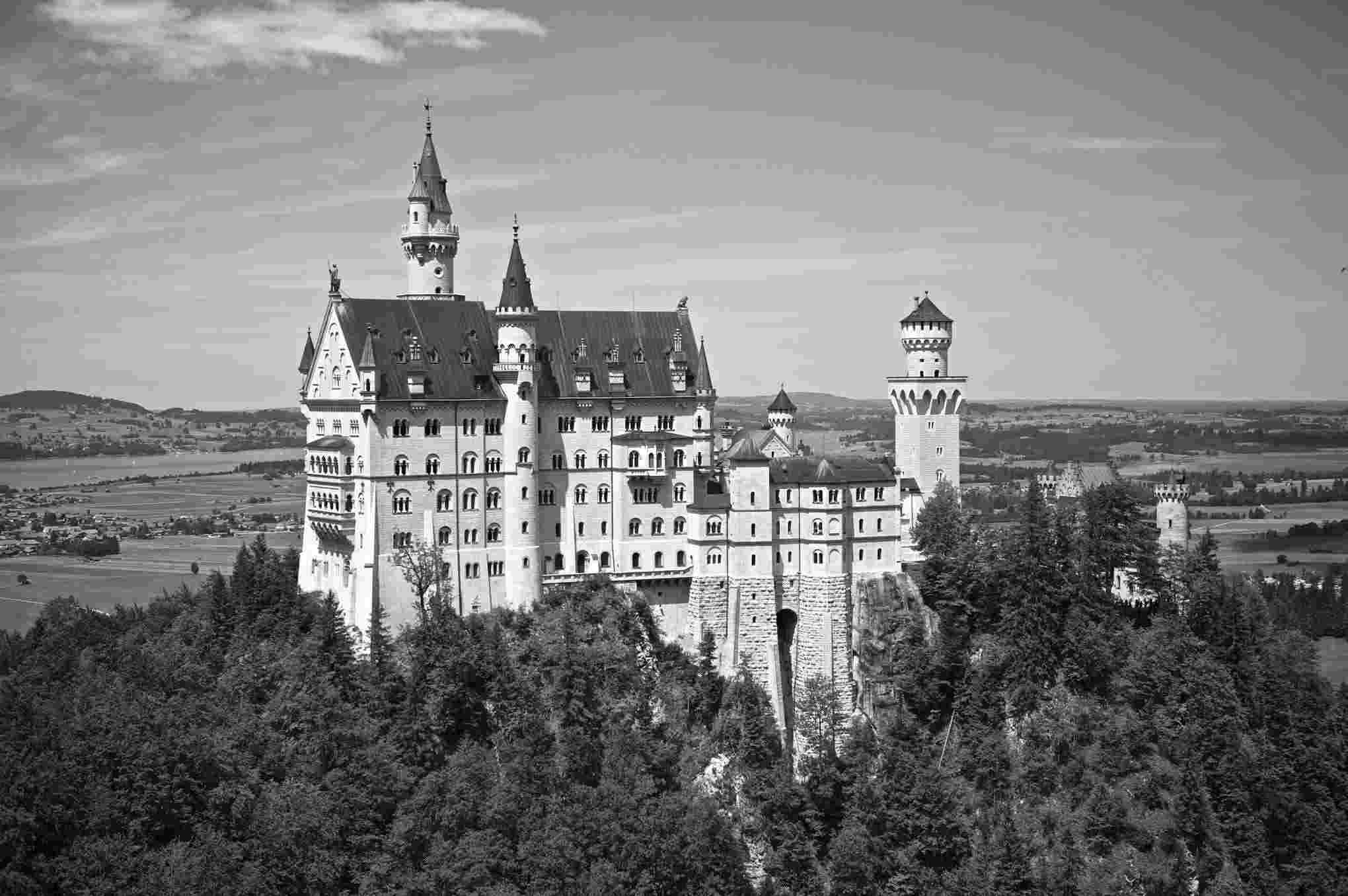}
    \includegraphics[width=\linewidth]{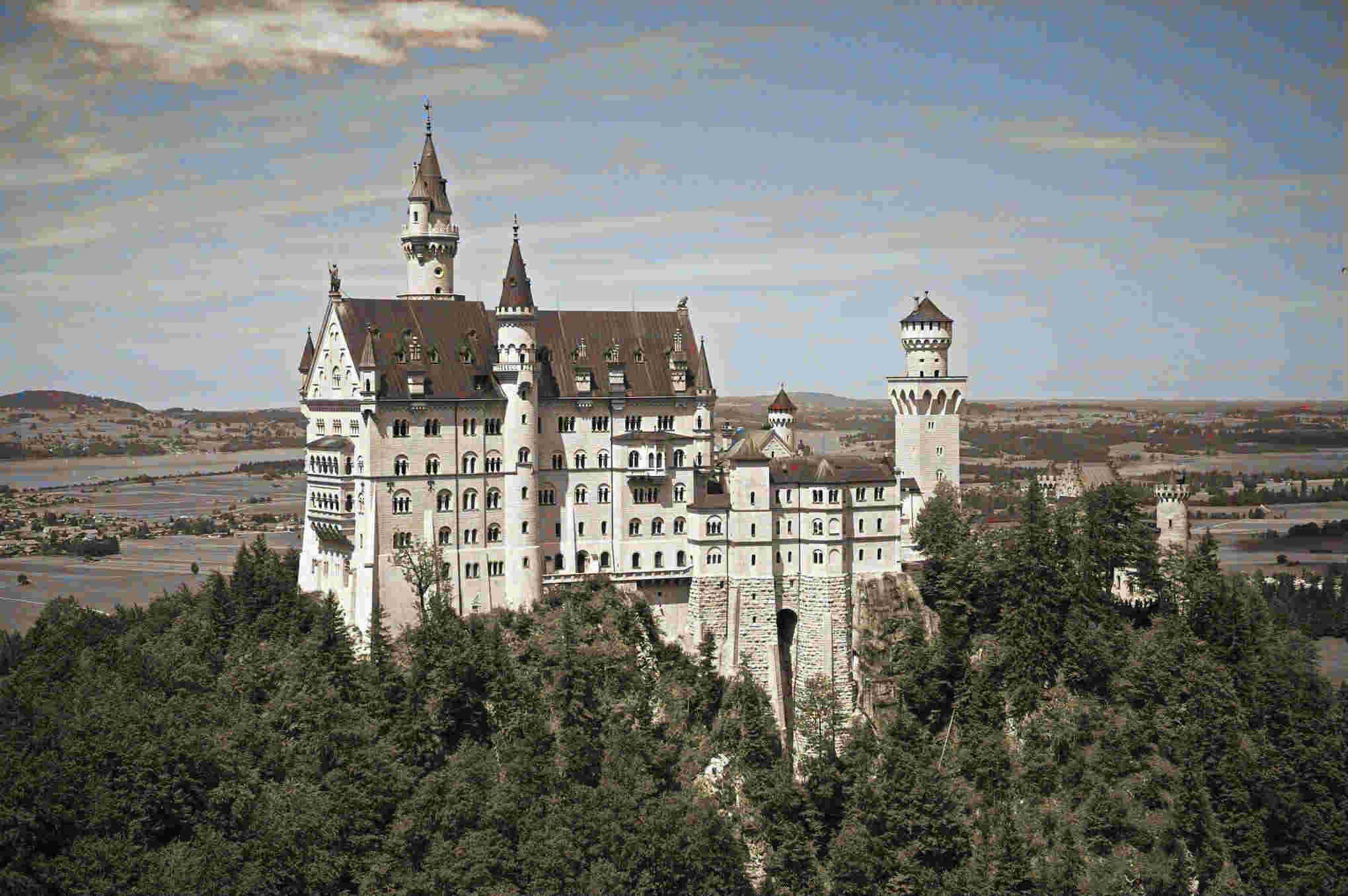}
    \caption{psnr=19.52,ssim=0.86}
    \label{fig:bad2}
  \end{subfigure}
  \begin{subfigure}[b]{0.20\linewidth}
    \includegraphics[width=\linewidth]{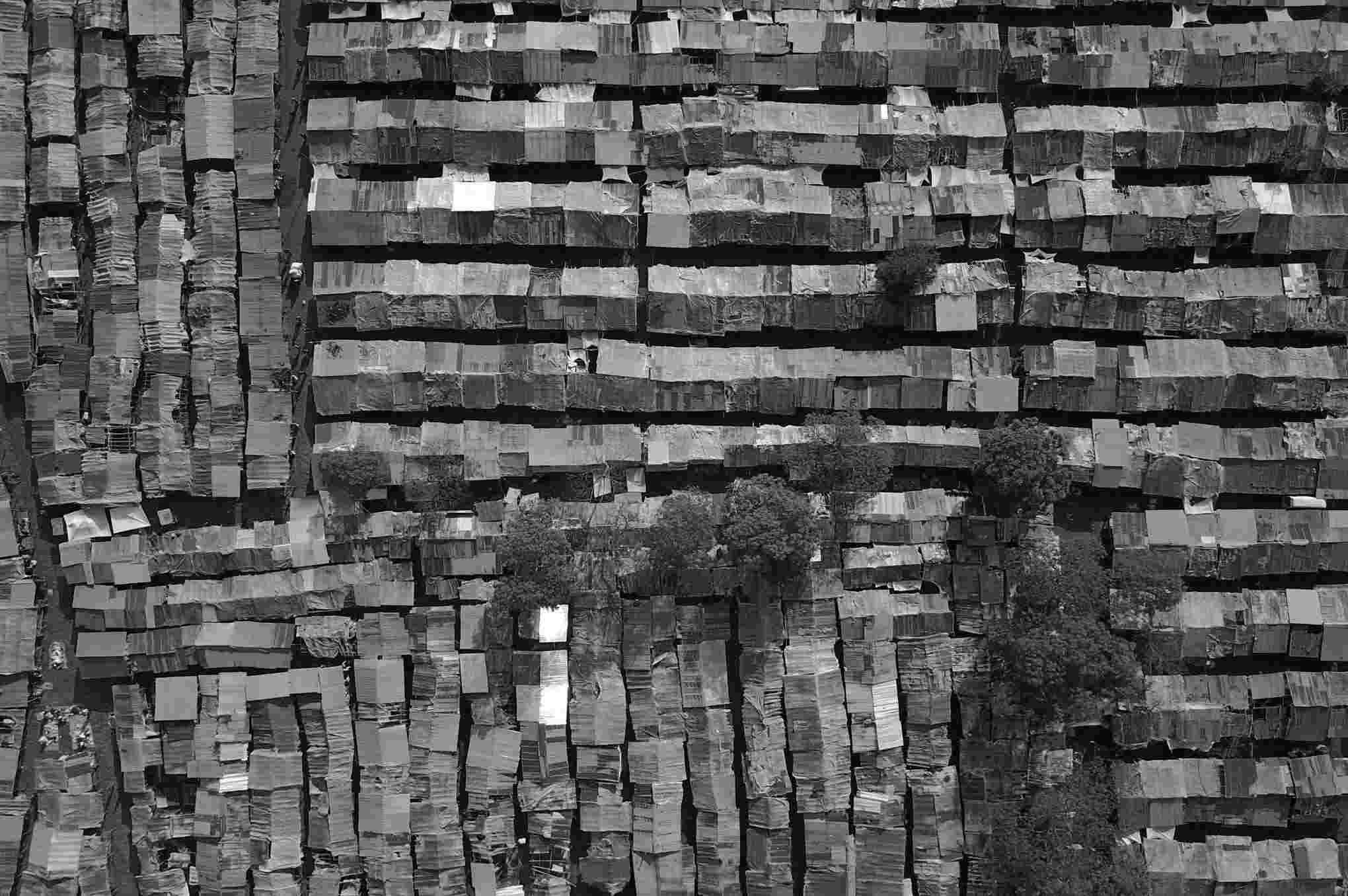}
    \includegraphics[width=\linewidth]{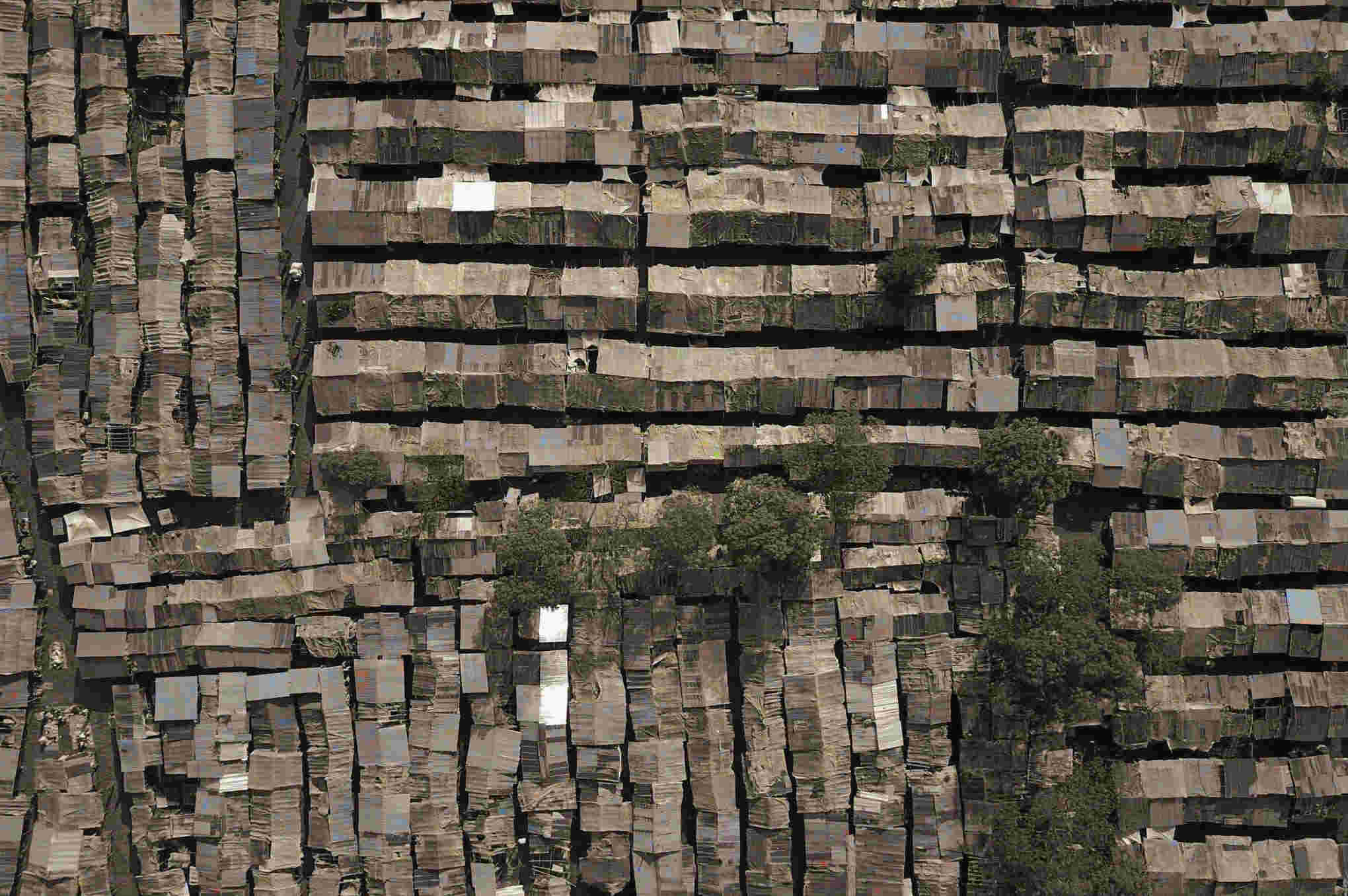}
    \caption{psnr=23.61,ssim=0.91}
  \end{subfigure}
  \begin{subfigure}[b]{0.20\linewidth}
    \includegraphics[width=\linewidth]{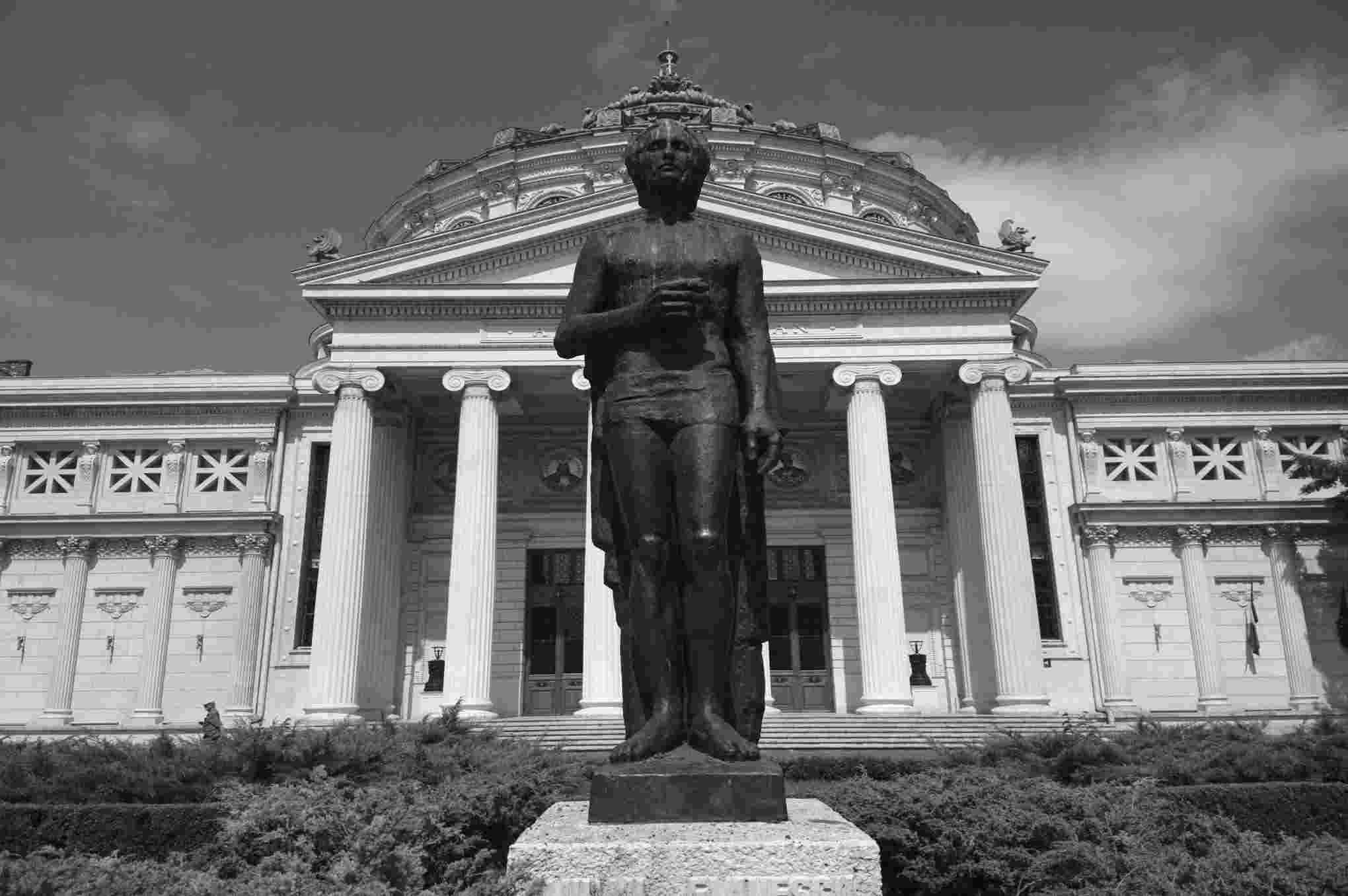}
    \includegraphics[width=\linewidth]{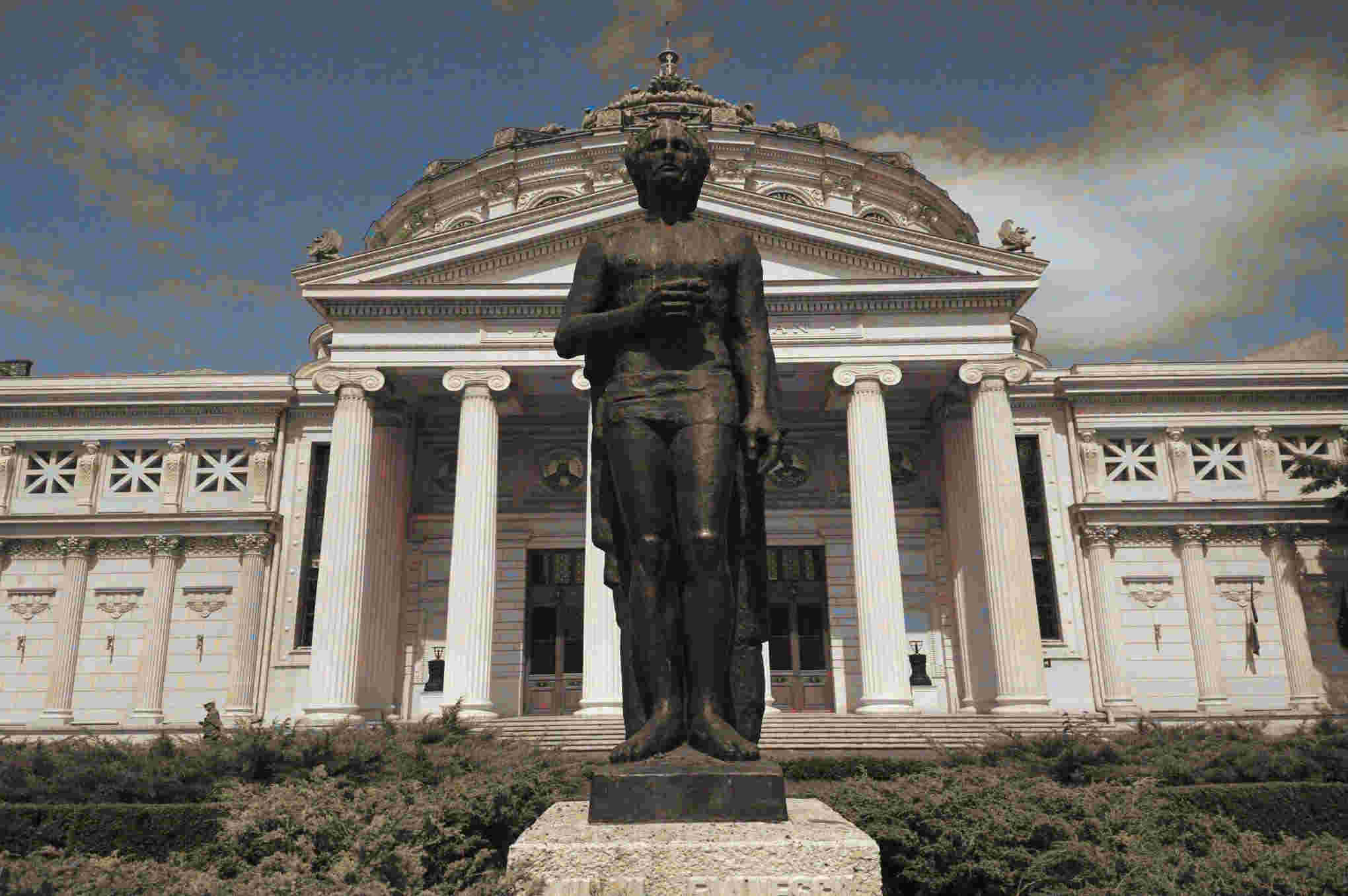}
    \caption{psnr=23.56,ssim=0.91}
  \end{subfigure}
  \begin{subfigure}[b]{0.20\linewidth}
    \includegraphics[width=\linewidth]{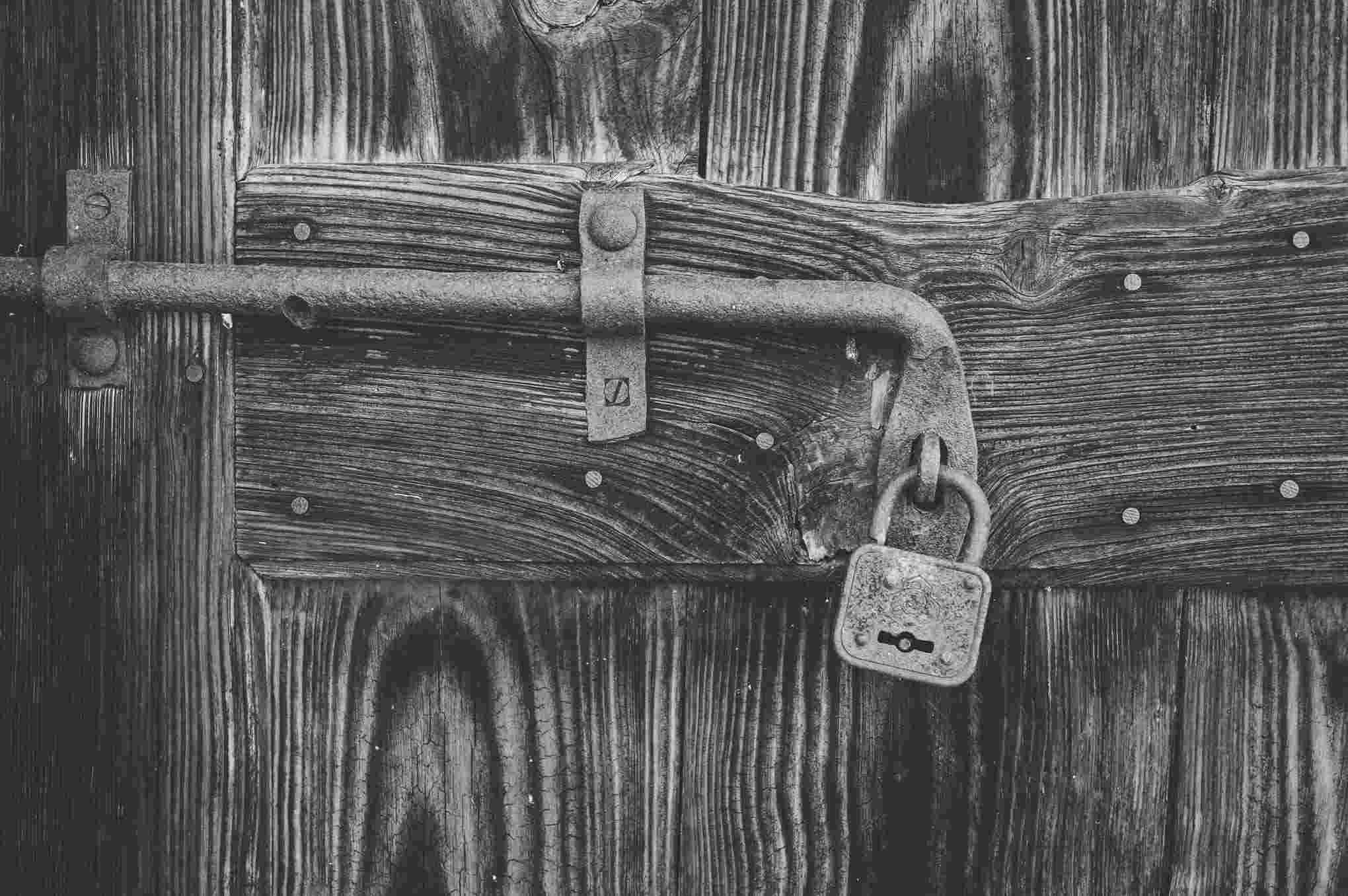}
    \includegraphics[width=\linewidth]{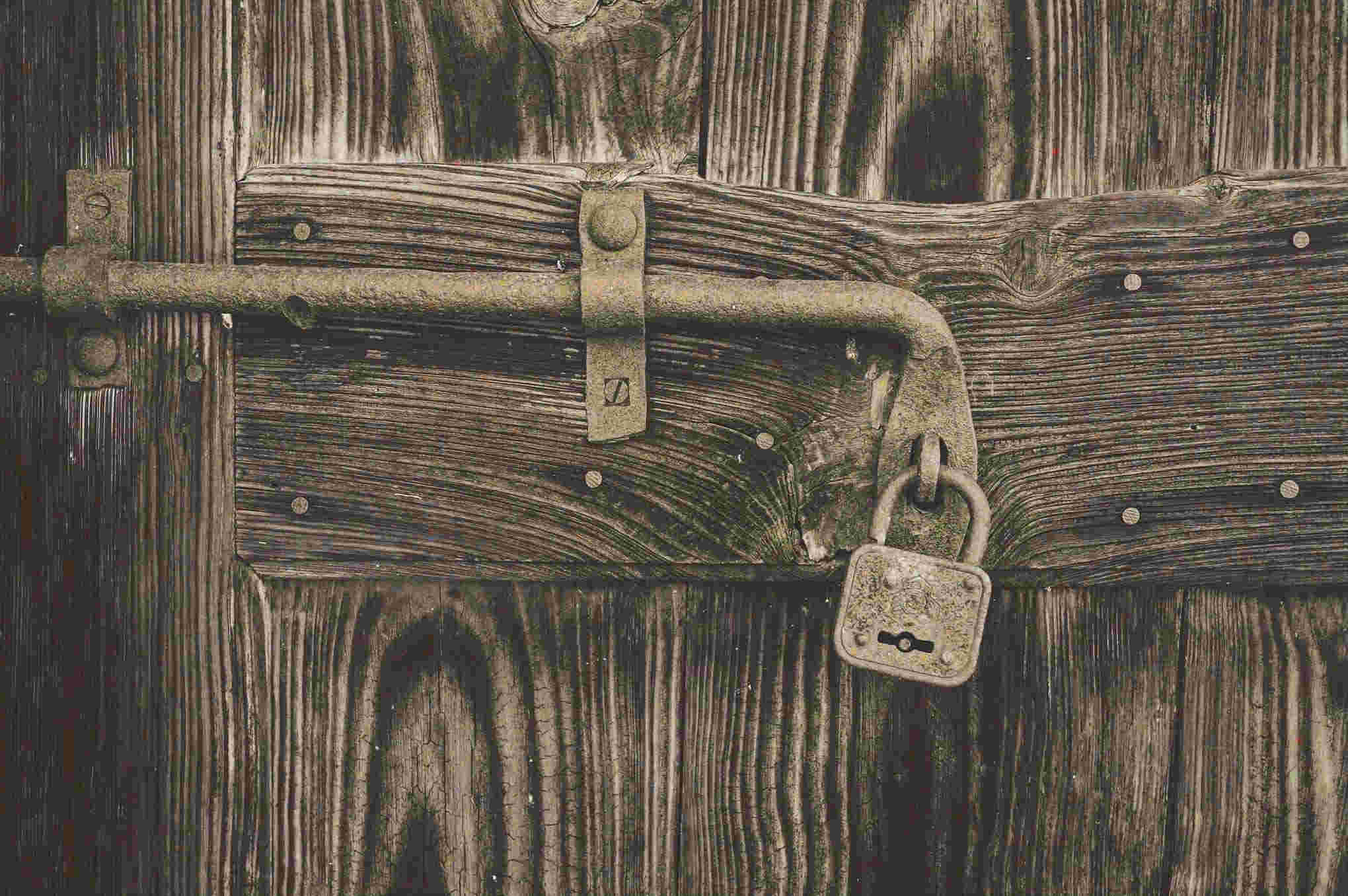}
    \caption{psnr=23.89,ssim=0.93}
  \end{subfigure}
  \begin{subfigure}[b]{0.20\linewidth}
    \includegraphics[width=\linewidth]{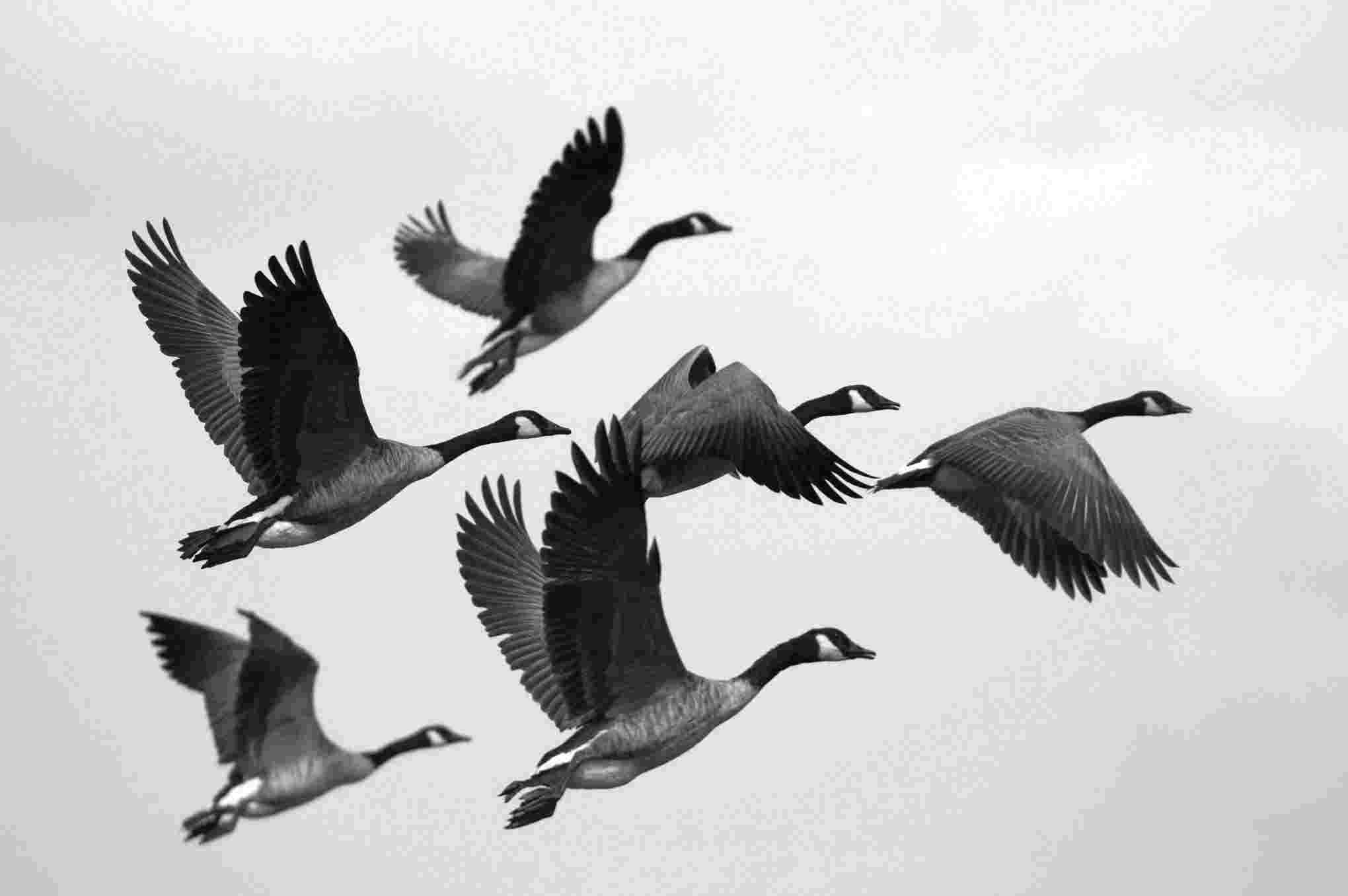}
    \includegraphics[width=\linewidth]{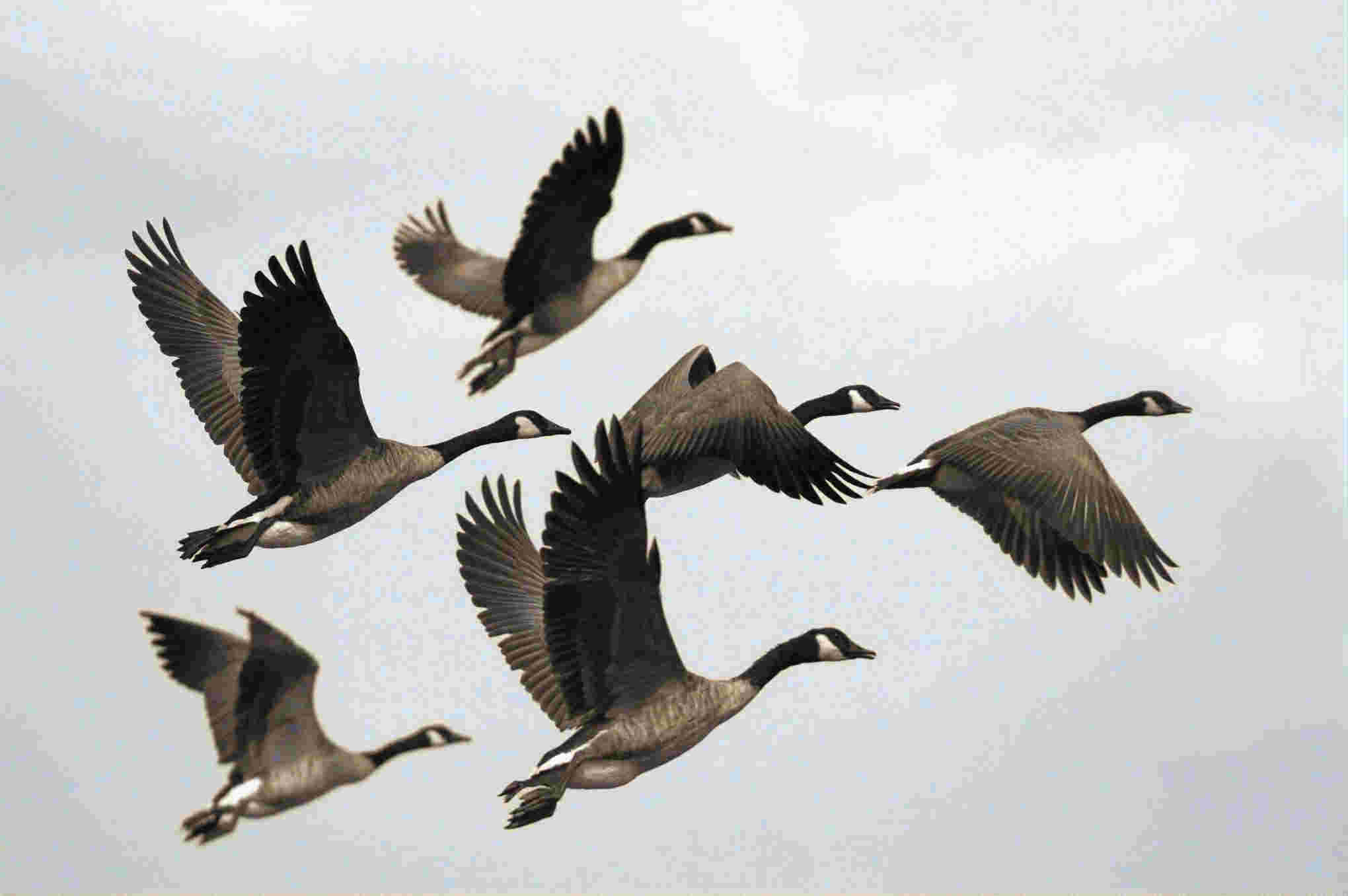}
    \caption{psnr=30.07,ssim=0.94}
  \end{subfigure}
  \begin{subfigure}[b]{0.20\linewidth}
    \includegraphics[width=\linewidth]{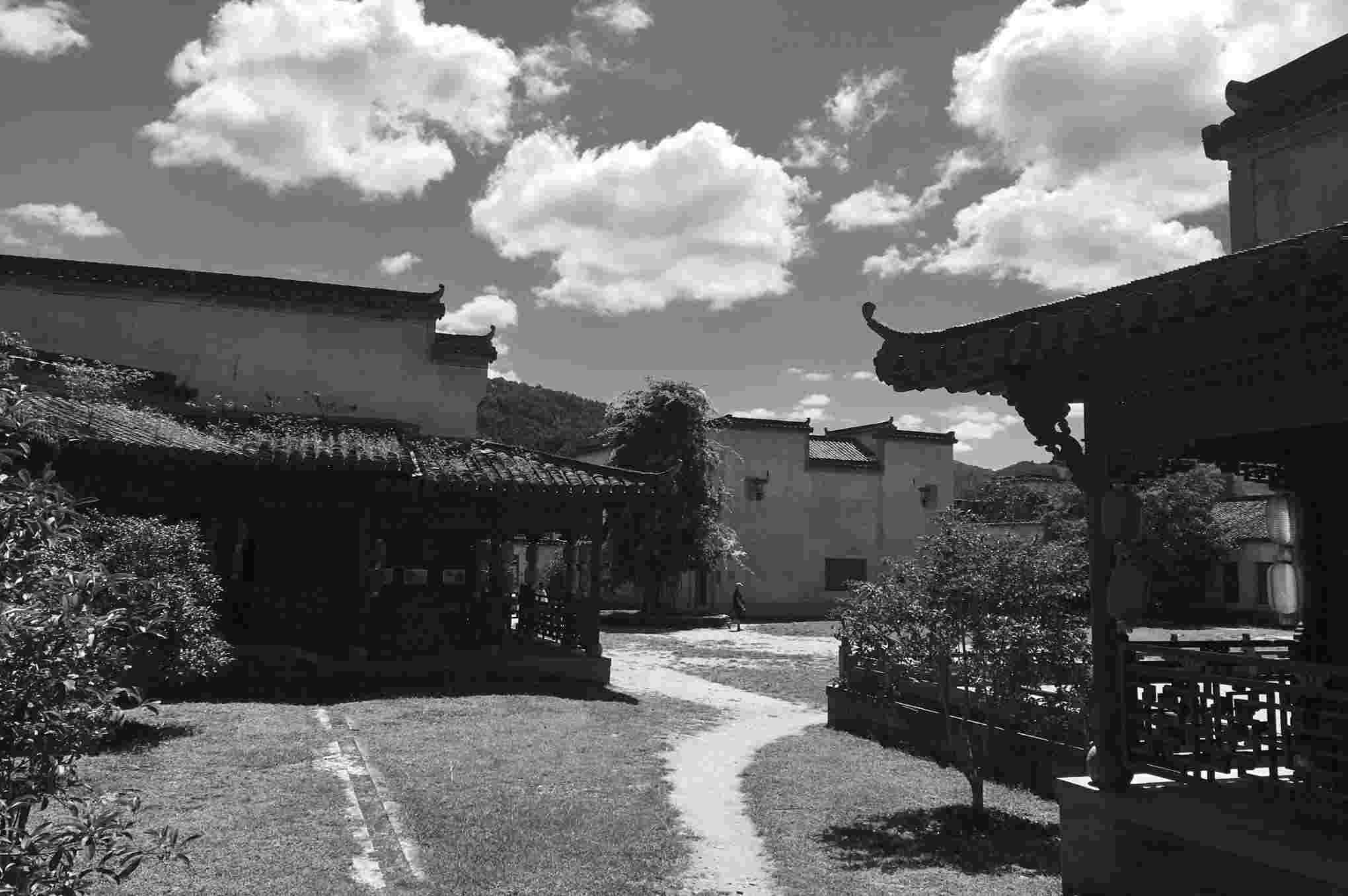}
    \includegraphics[width=\linewidth]{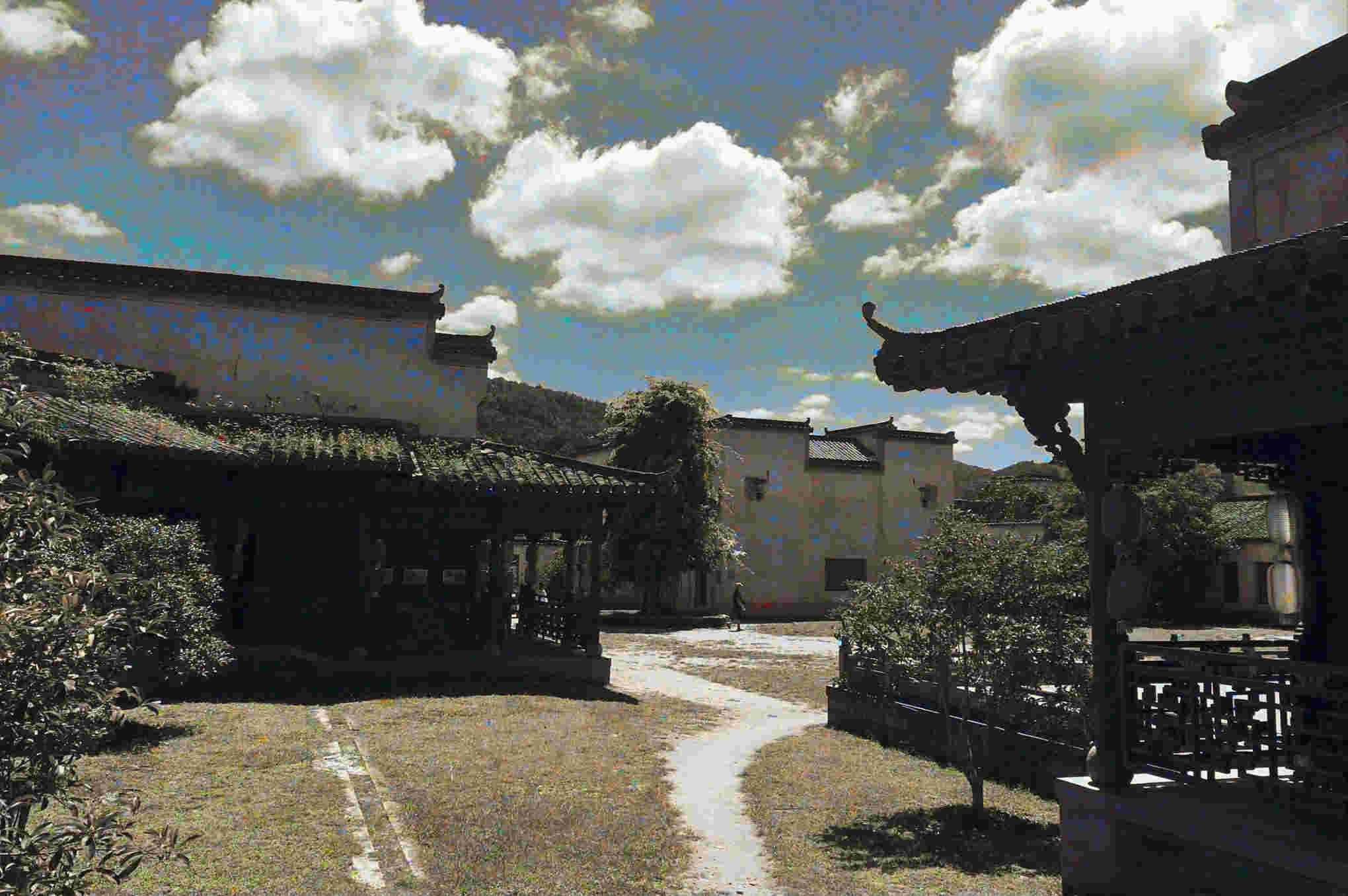}
    \caption{psnr=n/a,ssim=n/a}
  \end{subfigure}
  \begin{subfigure}[b]{0.20\linewidth}
    \includegraphics[width=\linewidth]{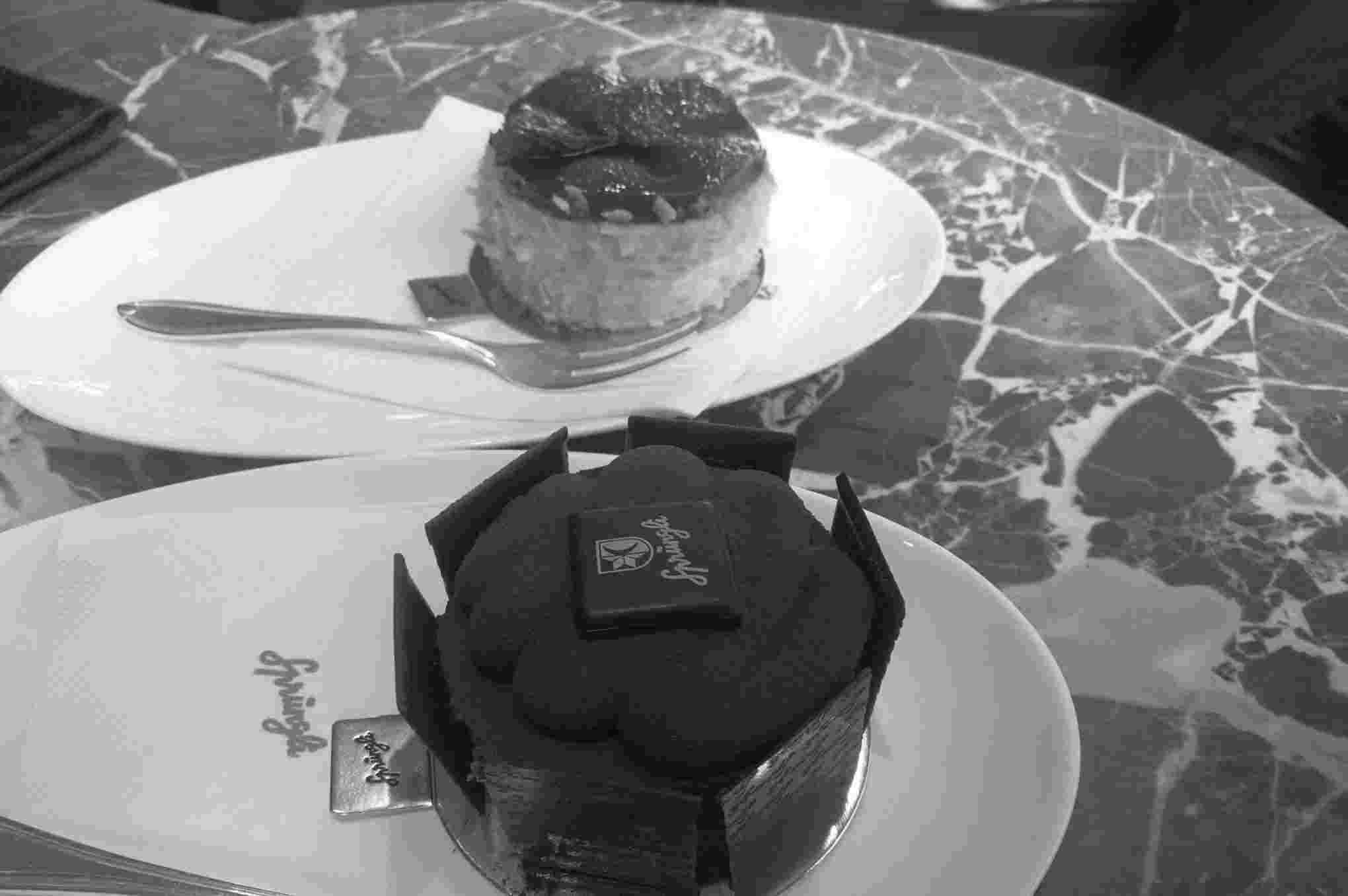}
    \includegraphics[width=\linewidth]{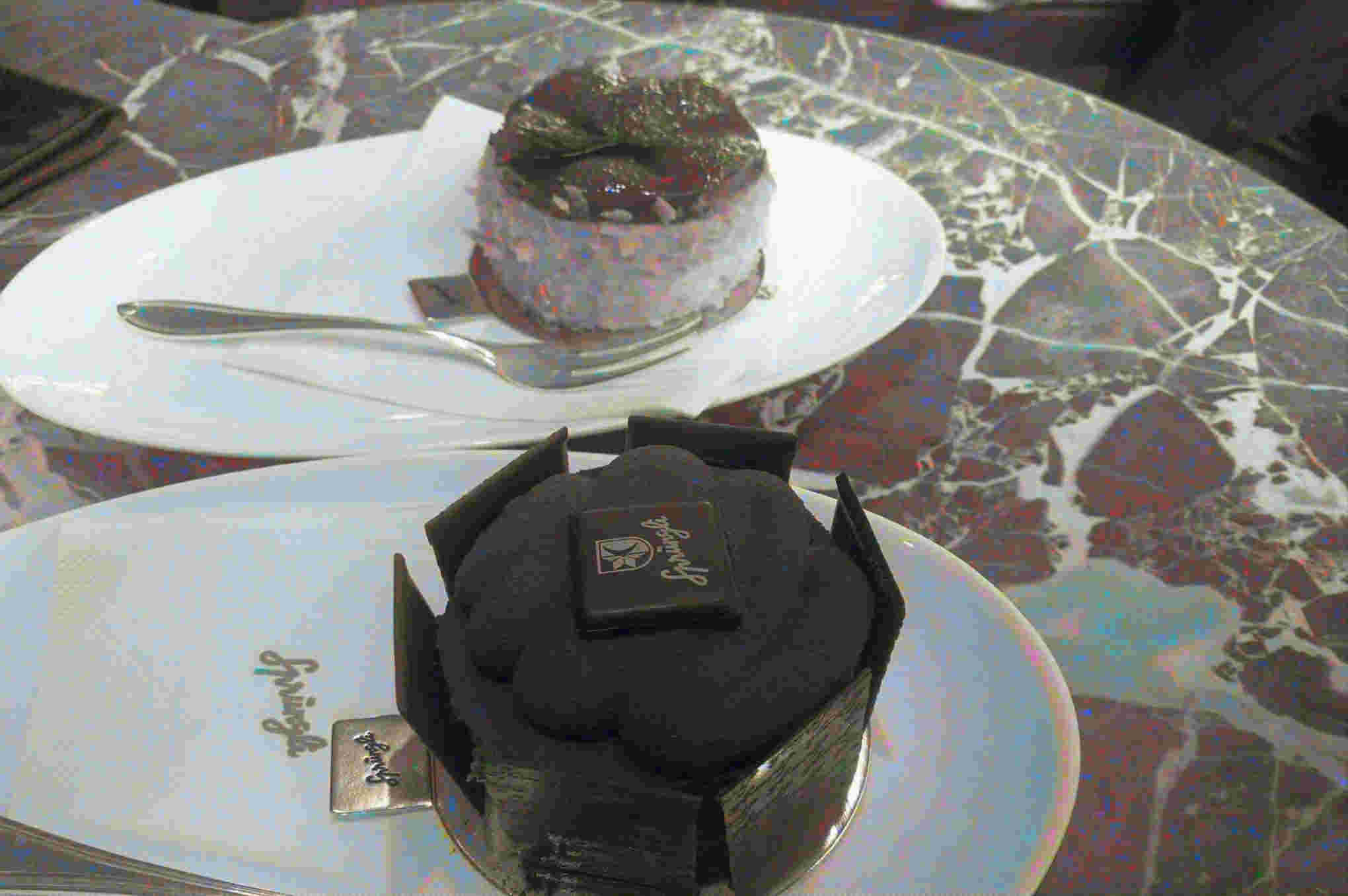}
    \caption{psnr=n/a,ssim=n/a}
  \end{subfigure}
  \caption{Some colorization results from DIV2K dataset.}
  \label{fig:result}
\end{figure*}

Some colorization results can be examined in Figure~\ref{fig:result}. According to the 
visual results, it can be said that the ColorCapsNet has promising colorization capability 
without any guidance. Even for some colorization results with low PSNR and SSIM values 
(Figure~\ref{fig:bad1} and Figure~\ref{fig:bad2}), it may generate satisfactory results. 
Those results in Figure~\ref{fig:result} with PSNR and SSIM values of 'n/a' indicate that 
there was no real color image corresponding to the estimated one and the evaluation 
metrics couldn't be calculated.

\begin{figure*}[h!]
  \centering
  \begin{subfigure}[b]{0.19\linewidth}
    \includegraphics[width=\linewidth]{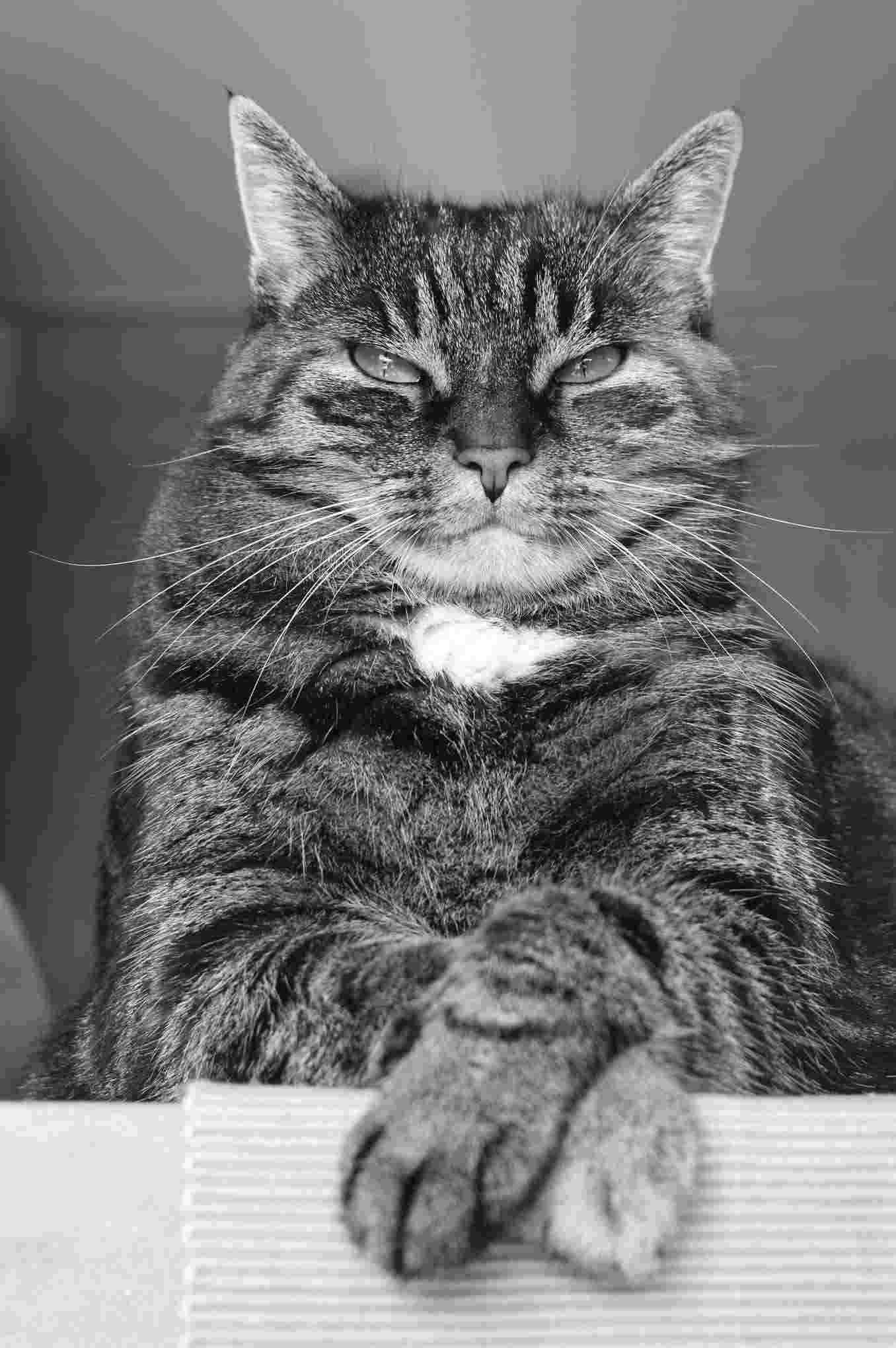}
    \caption{Grayscale}
  \end{subfigure}
  \begin{subfigure}[b]{0.19\linewidth}
    \centering{Epoch=5}
    \includegraphics[width=\linewidth]{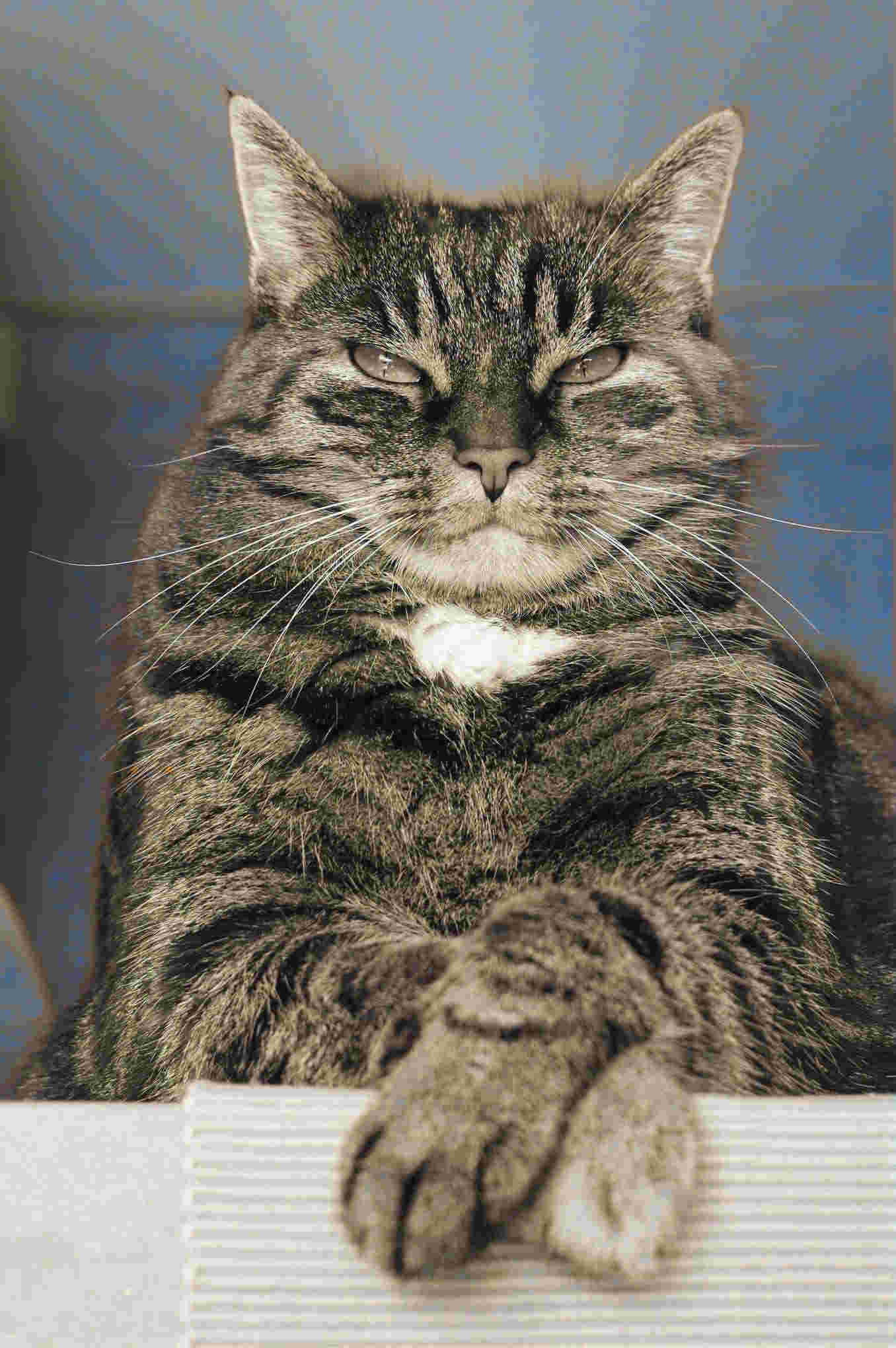}
    \caption{psnr=24.73,ssim=0.92}
  \end{subfigure}
  \begin{subfigure}[b]{0.19\linewidth}
    \centering{Epoch=10}
    \includegraphics[width=\linewidth]{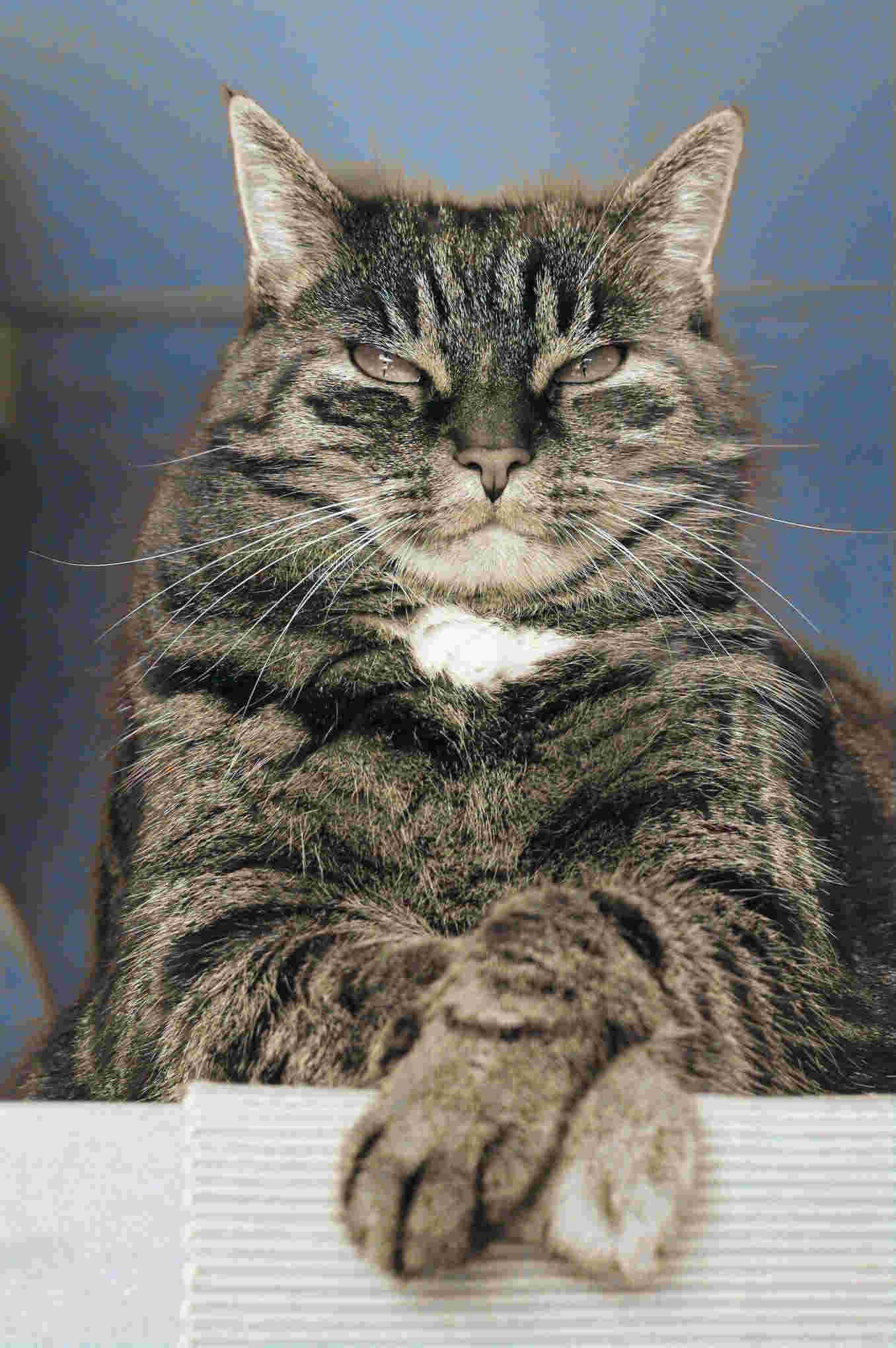}
    \caption{psnr=23.71,ssim=0.92}
  \end{subfigure}
  \begin{subfigure}[b]{0.19\linewidth}
    \centering{Epoch=15}
    \includegraphics[width=\linewidth]{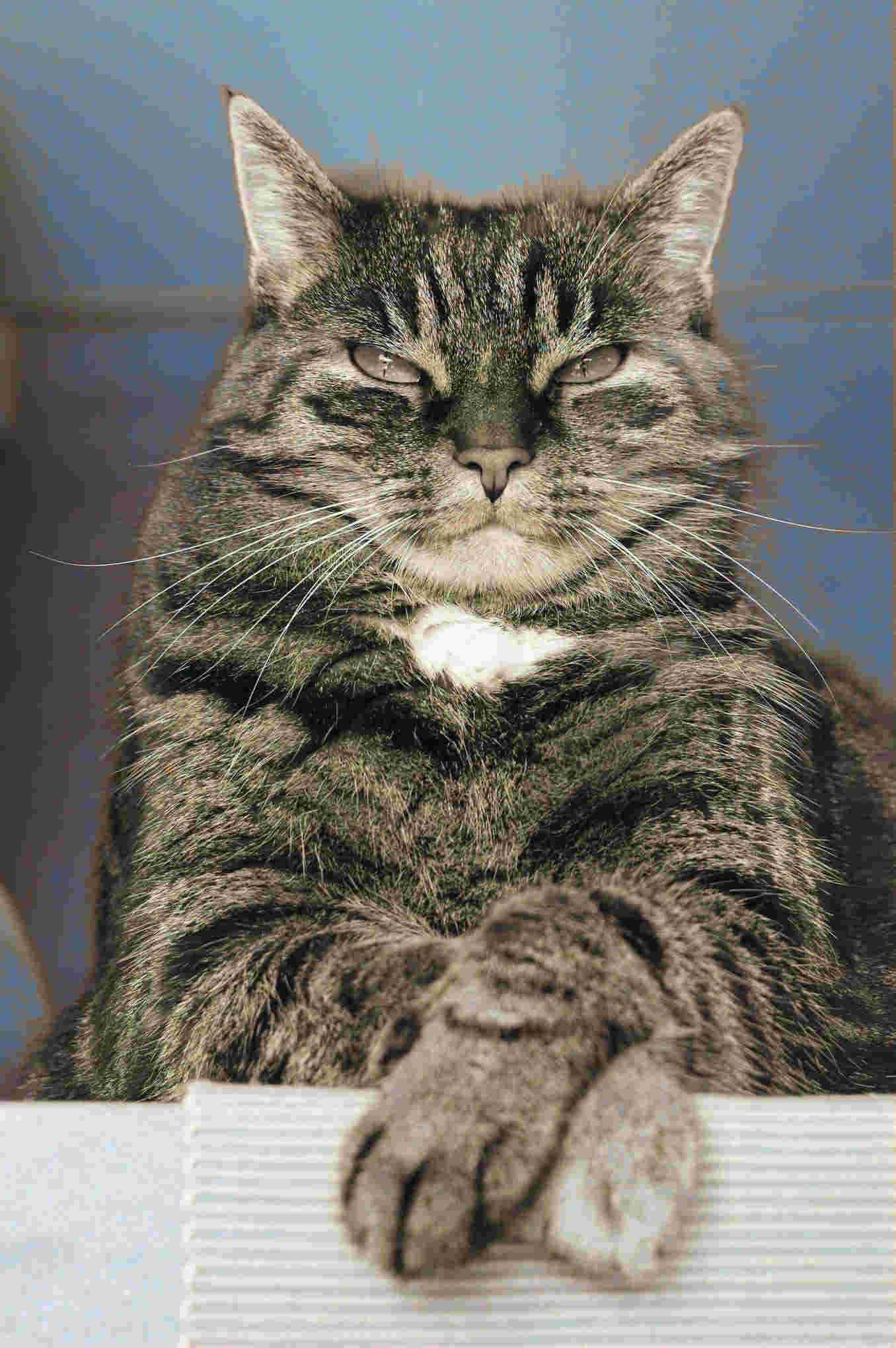}
    \caption{psnr=23.97,ssim=0.92}
  \end{subfigure}
  \begin{subfigure}[b]{0.19\linewidth}
    \centering{Epoch=50}
    \includegraphics[width=\linewidth]{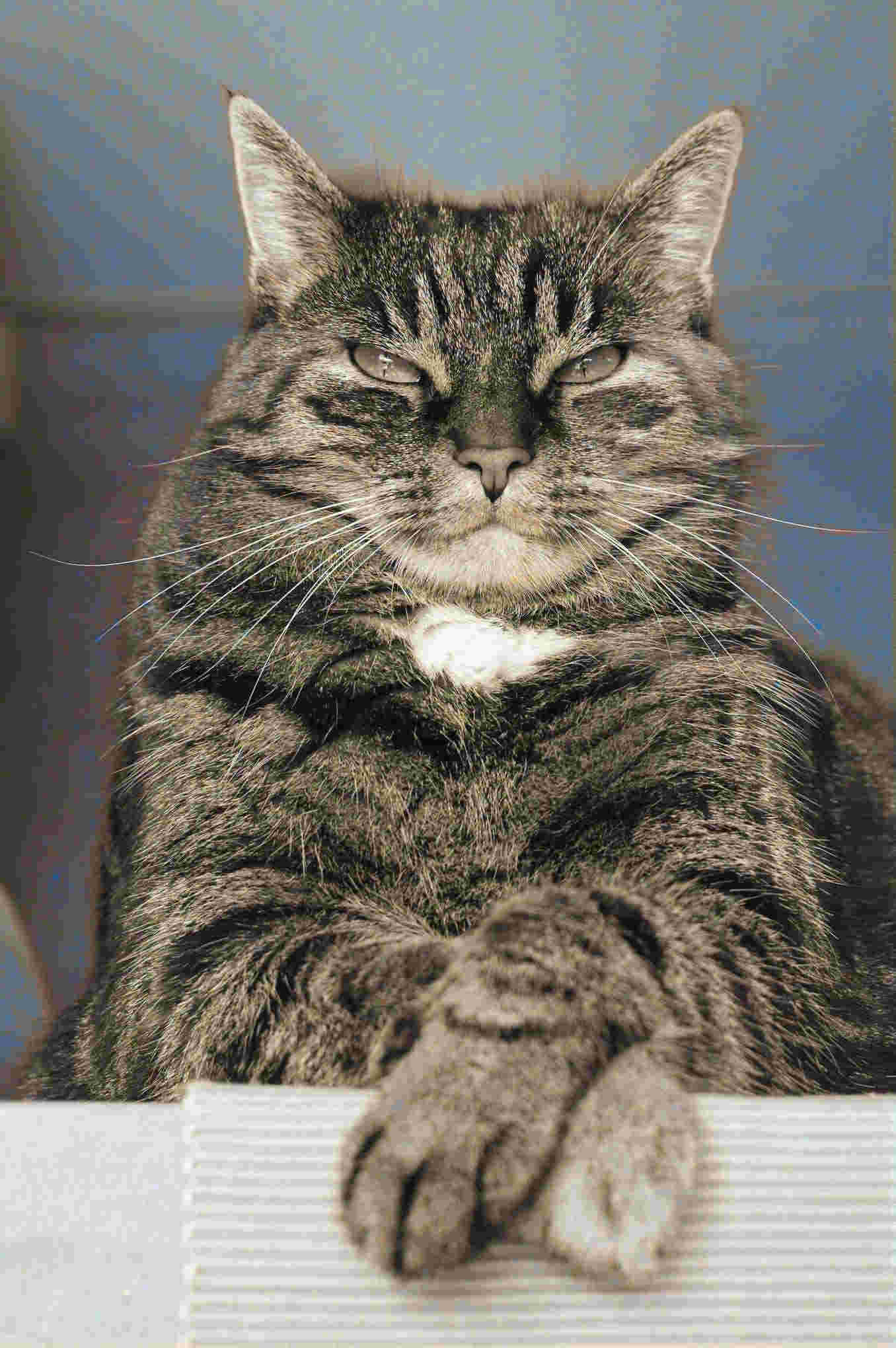}
    \caption{psnr=25.11,ssim=0.92}
  \end{subfigure}
  \begin{subfigure}[b]{0.19\linewidth}
    \includegraphics[width=\linewidth]{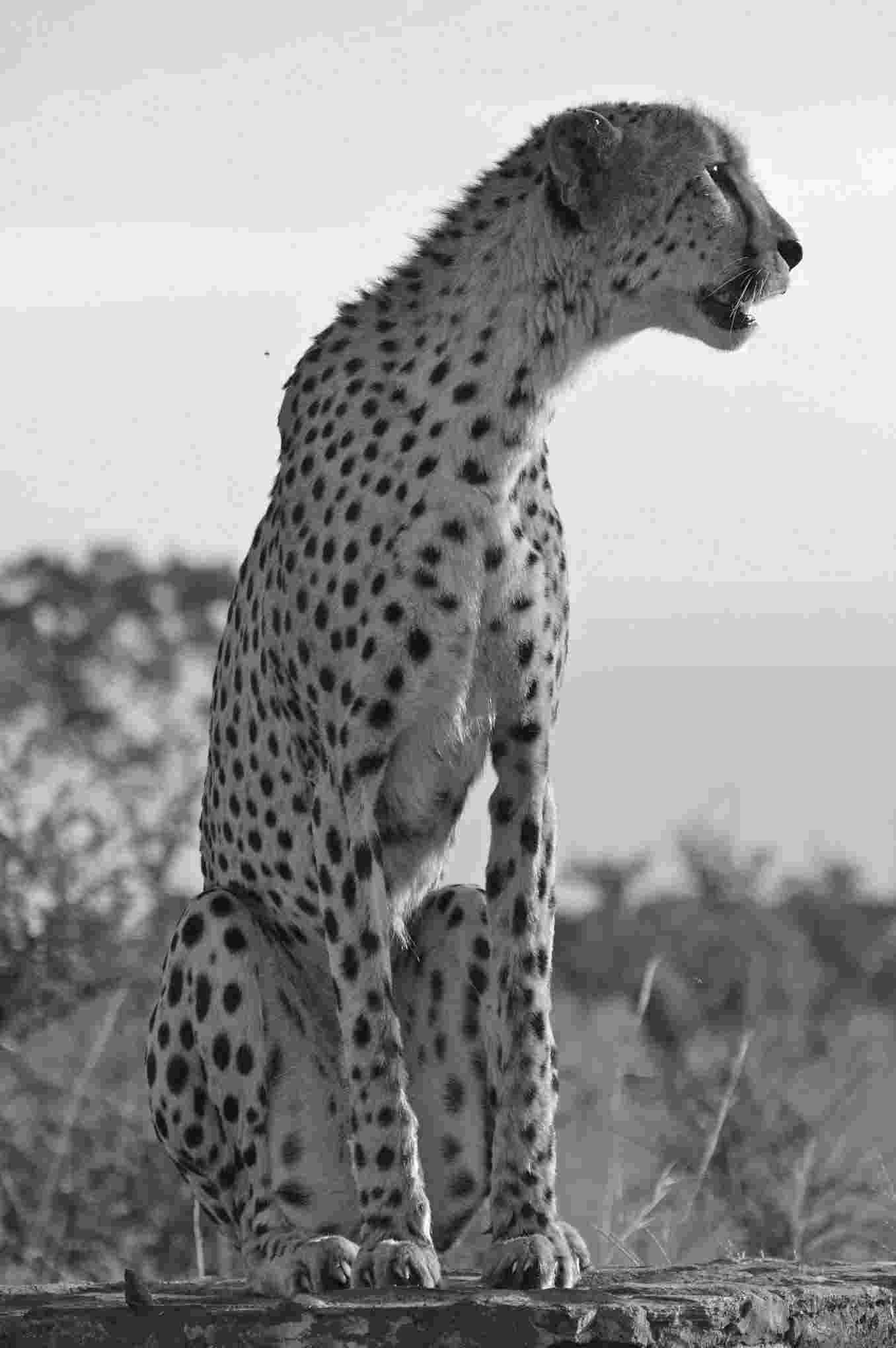}
    \caption{Grayscale}
  \end{subfigure}
  \begin{subfigure}[b]{0.19\linewidth}
    \includegraphics[width=\linewidth]{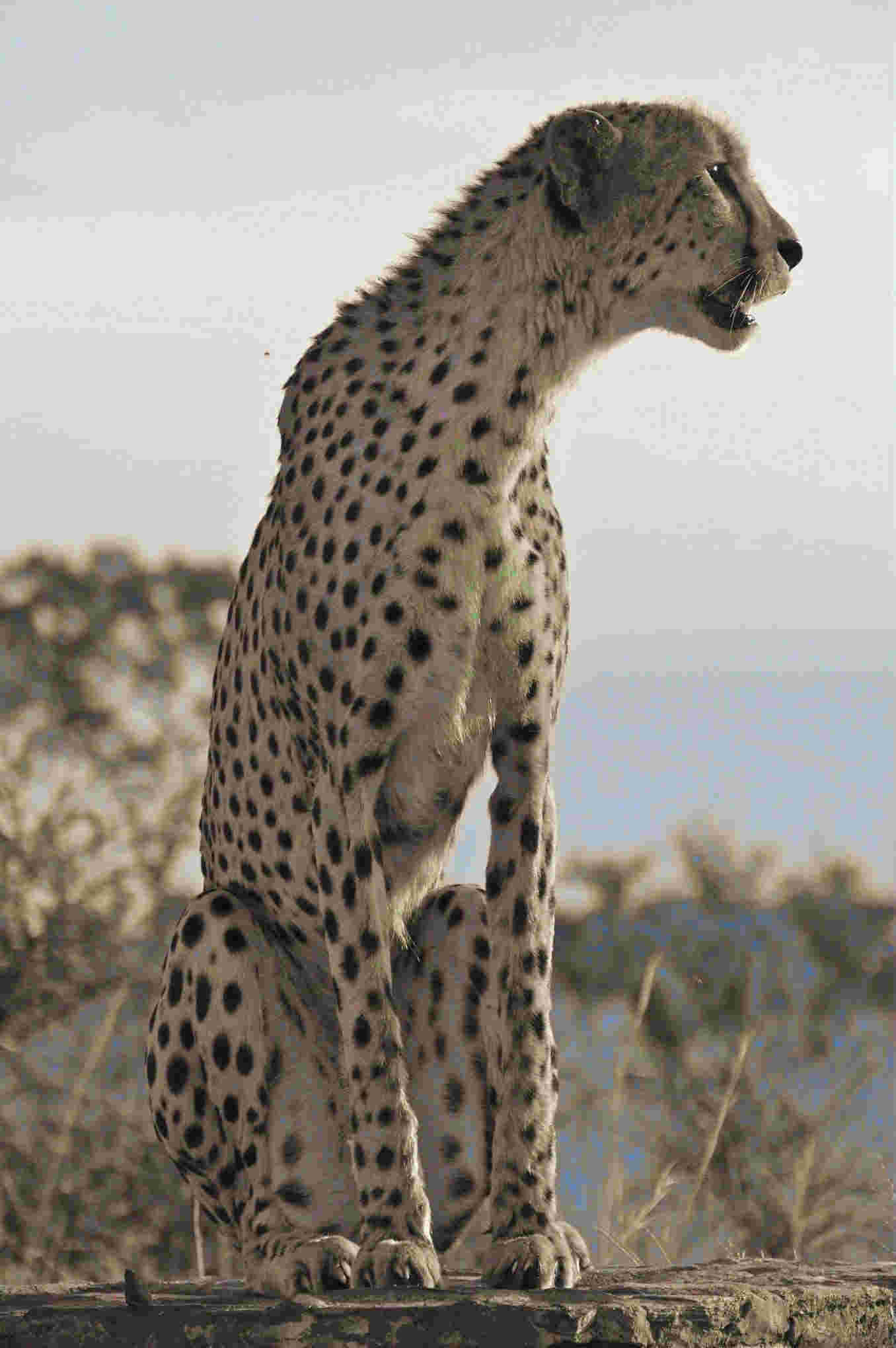}
    \caption{psnr=23.60,ssim=0.92}
  \end{subfigure}
  \begin{subfigure}[b]{0.19\linewidth}
    \includegraphics[width=\linewidth]{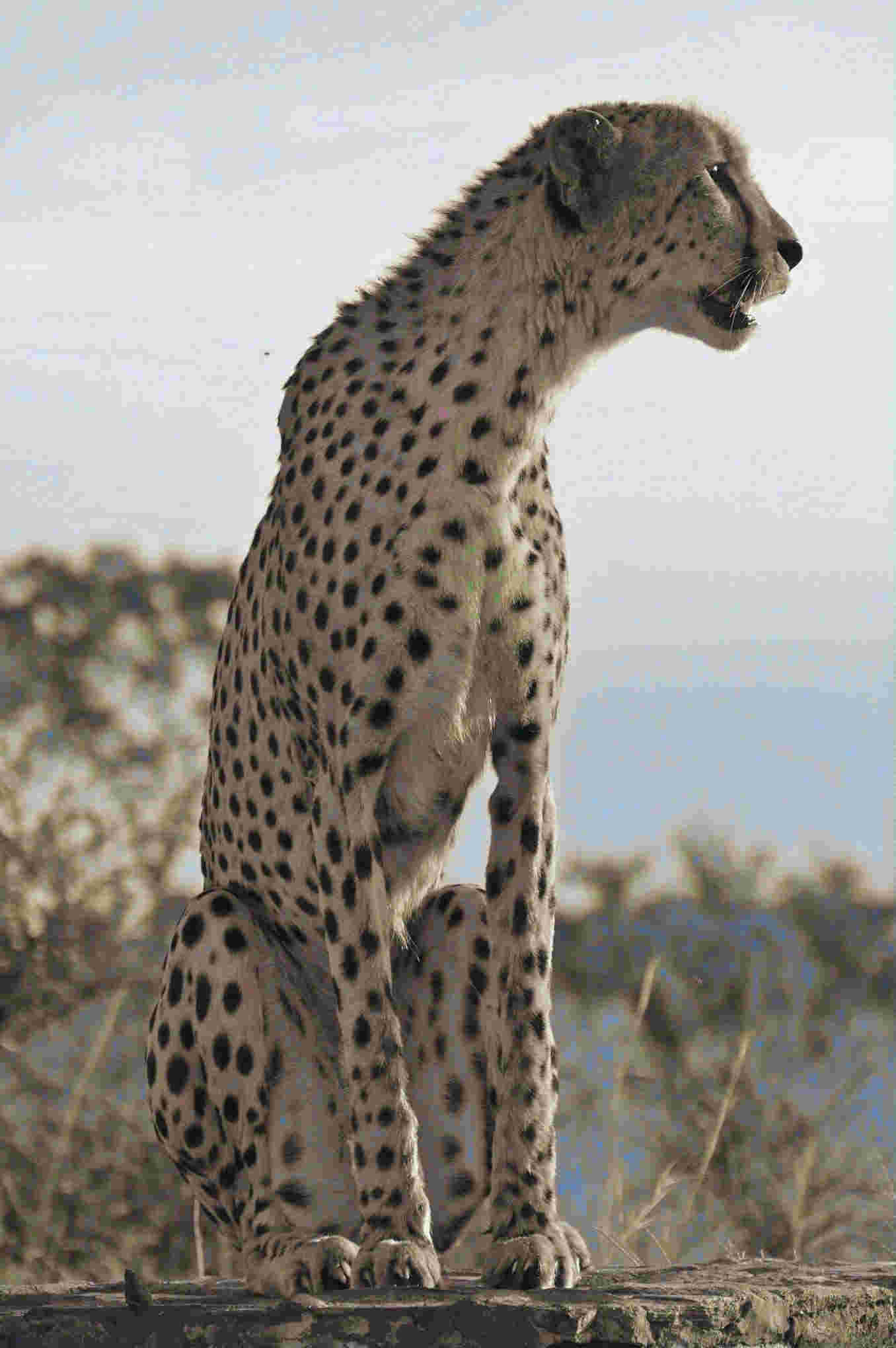}
    \caption{psnr=23.53,ssim=0.91}
  \end{subfigure}
  \begin{subfigure}[b]{0.19\linewidth}
    \includegraphics[width=\linewidth]{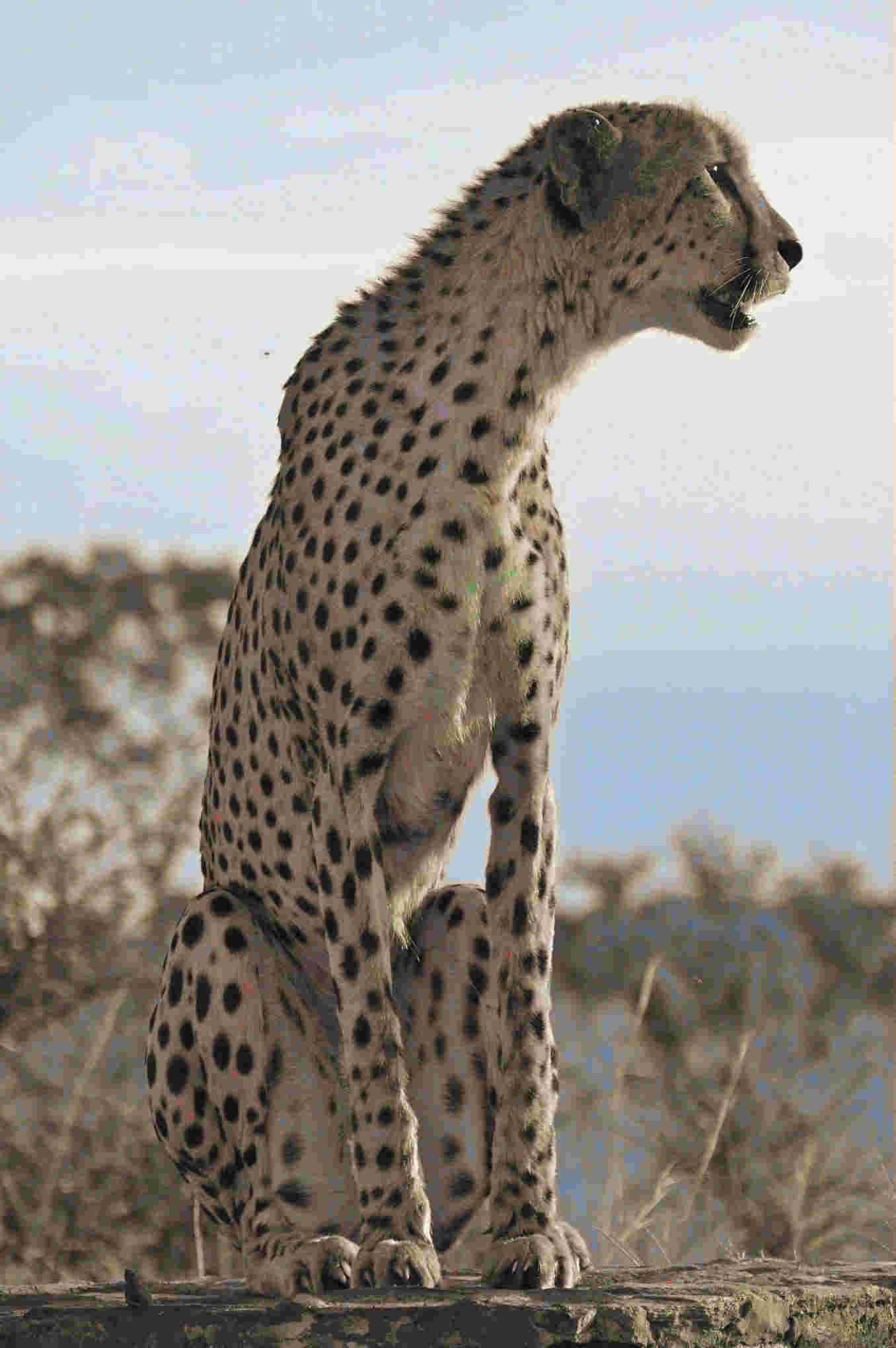}
    \caption{psnr=23.75,ssim=0.91}
  \end{subfigure}
  \begin{subfigure}[b]{0.19\linewidth}
    \includegraphics[width=\linewidth]{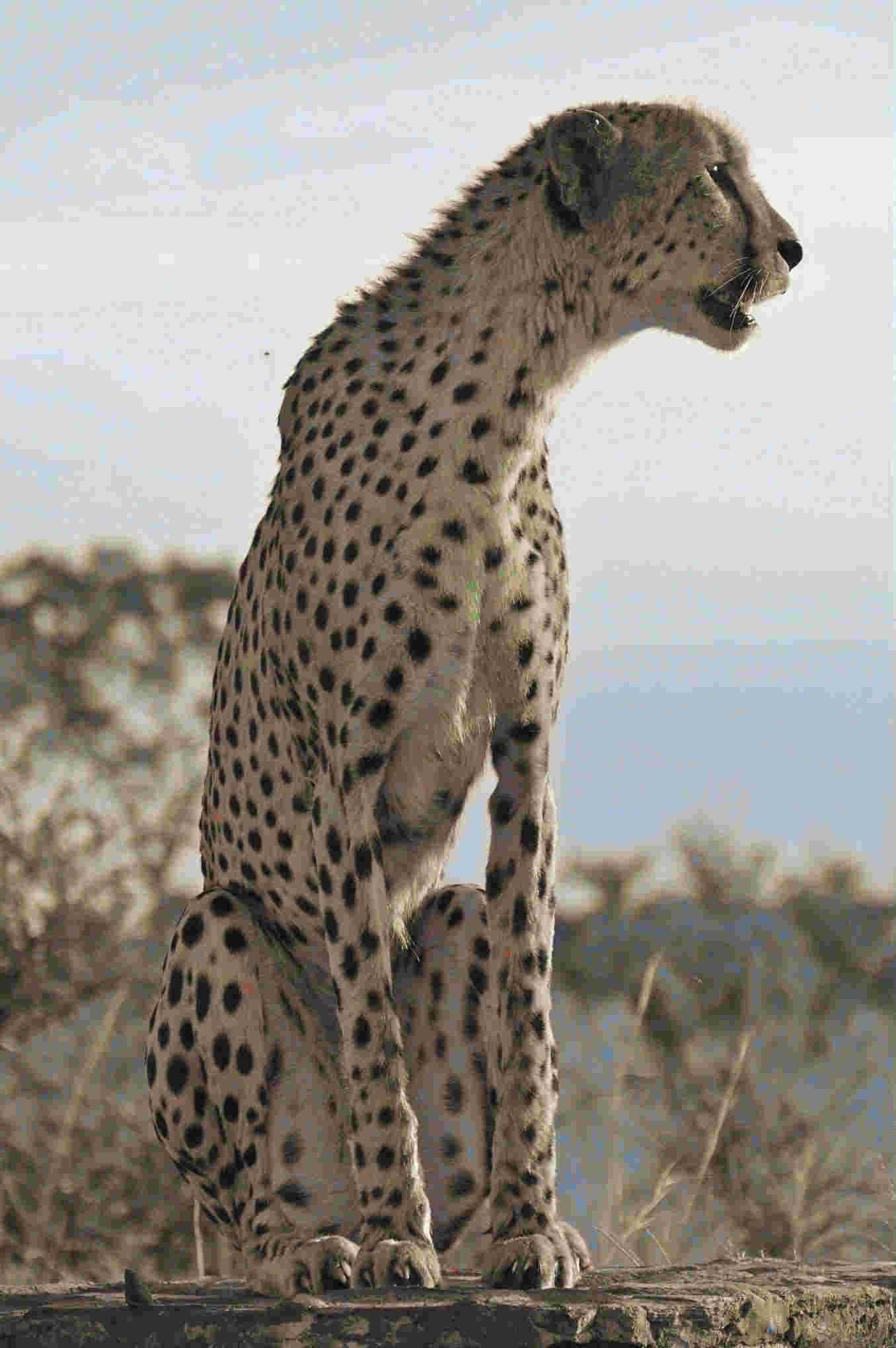}
    \caption{psnr=23.74,ssim=0.91}
  \end{subfigure}
  \caption{Colorization progress across training.}
  \label{fig:stepresult}
\end{figure*}

In Figure~\ref{fig:stepresult}, the progress in the colorization during training can be examined for two 
instances from the DIV2K validation set. As seen in the figure, although the PSNR values 
oscillate softly with the changing number of the epoch ($24.38 \pm 0.56$ for the top row and $23.65 \pm 0.09$ for the bottom row), the SSIM values are stabilized as indicating 
no more improvement on the colorization performance. The effect of the longer training on the colorization 
performance should be further investigated in order to see if it helps to the improvement as in \cite{sabour2017dynamic}.

\section{Conclusion} \label{conclusion}

In this paper, a simple Capsule Network (CapsNet) architecture is designed for ill-posed 
image colorization problem. Proposed method is a fully automatic, end-to-end, patch-based 
deep model named as Colorizer Capsule Network (ColorCapsNet) and it exploits the generative 
and segmentation capabilities of the CapsNets for the colorization task. Experiments show that 
the ColorCapsNet has promising and comparable results on provided DIV2K dataset with its 
simple design and deserves further investigation such as designing deeper CapsNet architectures or 
integrating multimodal color distribution for better colorization performance.

{\small
\bibliographystyle{ieee_fullname}
\bibliography{capsule}
}

\end{document}